\documentclass[usegraphicx,usenatbib]{mn2e}
\usepackage{amsmath}
\usepackage{amssymb}
\usepackage{times}
\usepackage{subfigure}
\usepackage{epsfig}

\renewcommand{\descriptionlabel}[1]%
 {\hspace{\labelsep}\textbf{#1}}

\title[Light and Motion in SDSS Stripe 82]
  {Light and Motion in SDSS Stripe 82: The Catalogues}

\author[D.M. Bramich et al.]
  {D.M.~Bramich$^{1,2}$\thanks{E-mail: dmb@ing.iac.es},
   S.~Vidrih$^{1,3}$,
   L.~Wyrzykowski$^1$,
   J.A.~Munn$^4$,
   H.~Lin$^5$,
   N.W.~Evans$^1$,
   \newauthor
   M.C.~Smith$^1$,
   V.~Belokurov$^1$,
   G.~Gilmore$^1$,
   D.B.~Zucker$^1$,
   P.C.~Hewett$^1$,
   L.L.~Watkins$^1$,
   \newauthor
   D.C.~Faria$^1$,
   M.~Fellhauer$^1$,
   G.~Miknaitis$^5$,
   D.~Bizyaev$^6$,
   \v{Z}.~Ivezi\'{c}$^7$,
   \newauthor
   D.P.~Schneider$^8$,
   S.A.~Snedden$^6$,
   E.~Malanushenko$^6$,
   V.~Malanushenko$^6$,
   K.~Pan$^6$
   \newauthor   
 \medskip
 \\$^1$Institute of Astronomy, University of Cambridge, Madingley Road,
       Cambridge, CB3 0HA, UK
 \\$^2$Isaac Newton Group of Telescopes, Apartado de Correos 321,
       E-38700 Santa Cruz de la Palma, Canary Islands, Spain
 \\$^3$Astronomisches Rechen-Institut/Zentrum f\"ur Astronomie der 
  Universit\"at  Heidelberg, M\"onchhofstrasse 12-14, 69120 Heidelberg, 
  Germany
 \\$^4$US Naval Observatory, Flagstaff Station, 10391 W. Naval Observatory Road, 
       Flagstaff, AZ 86001-8521, USA
 \\$^5$Fermi National Accelerator Laboratory, Box 500, Batavia, IL 60510, USA
 \\$^6$Apache Point Observatory, 2001 Apache Point Rd., Sunspot, NM 88349, USA
 \\$^7$Department of Astronomy, University of Washington, Seattle, WA 98155, USA
 \\$^8$Department of Astronomy and Astrophysics, The Pennsylvania State University, University Park, PA 16802, USA
    }

\begin{document}

\date{Accepted 2007 August ???. Received 2007 August ???; Submitted 2007 August ???}

\pagerange{\pageref{firstpage}--\pageref{lastpage}} \pubyear{2007}

\maketitle

\label{firstpage}

\begin{abstract} 
We present a new public archive of light-motion curves in Sloan
Digital Sky Survey (SDSS) Stripe 82, covering 99$^{\circ}$ in right ascension
from $\alpha = 20.7^{\mbox{\small h}}$ to 3.3$^{\mbox{\small h}}$ and spanning
2\fdg52 in declination from $\delta = -$1\fdg26 to 1\fdg26, for a total sky area of $\sim$249~deg$^{2}$. 
Stripe 82 has been repeatedly monitored in the $u$, $g$, $r$, $i$ and $z$ bands over a seven-year
baseline. Objects are cross-matched between runs, taking into account the 
effects of any proper motion. The resulting catalogue contains almost 4~million
light-motion curves of stellar objects and galaxies. The photometry 
are recalibrated to correct for varying photometric zeropoints, achieving 
$\sim$20~mmag and $\sim$30~mmag root-mean-square (RMS) accuracy down to 18~mag 
in the $g$, $r$, $i$ and $z$ bands for point sources and extended sources, respectively.
The astrometry are recalibrated to correct for inherent systematic errors in the
SDSS astrometric solutions, achieving $\sim$32~mas and $\sim$35~mas RMS accuracy down to 18~mag
for point sources and extended sources, respectively.

For each light-motion curve, 229 photometric and astrometric quantities are
derived and stored in a higher-level catalogue. On the photometric
side, these include mean exponential and PSF magnitudes along with uncertainties, 
RMS scatter, $\chi^2$ per degree of freedom, various magnitude
distribution percentiles, object type (stellar or galaxy), and
eclipse, Stetson and Vidrih variability indices.
On the astrometric side, these quantities include mean positions, proper motions
as well as their uncertainties and $\chi^2$ per degree of freedom. 
The here presented light-motion curve
catalogue is complete down to $r~\sim~21.5$ and is at present the deepest 
large-area photometric and astrometric variability catalogue available.
\end{abstract} 


\section{Introduction}

\begin{table*}
\centering
\caption{The list of SDSS imaging runs included in the LMCC. All runs were processed with version 40 of the SDSS {\tt frames} pipeline
         except those runs marked with an asterisk, which were processed with version 41. Runs 4203 and 5823 (underlined) are the reference
         runs used for the astrometric calibrations (see Section~2.4).}
{\scriptsize
\begin{tabular}{@{}llll}
\hline
Year & Month & North Strip Runs & South Strip Runs \\
\hline
1998 & Sep & 94 & 125 \\
 & & & \\
1999 & Oct & 1033 & 1056 \\
 & & & \\
2000 & Sep & 1752 & - \\
2000 & Oct & - & 1755 \\
2000 & Nov & - & 1894 \\
 & & & \\
2001 & Jun & 2385 & - \\
2001 & Sep & 2570, 2578, 2589 & 2579, 2583, 2585 \\
2001 & Oct & 2649, 2650, 2659, 2662, 2677 & - \\
2001 & Nov & 2700, 2708, 2728, 2738  & 2709 \\
2001 & Dec & 2768, 2820 & - \\
2002 & Jan & 2855, 2861, 2873 & 2886 \\
 & & & \\
2002 & Sep & - & 3325$^{*}$ \\
2002 & Oct & 3362, 3384, 3437 & 3355, 3360$^{*}$, 3388, 3427, 3430, 3434, 3438 \\
2002 & Nov & 3461 & 3460, 3465 \\
 & & & \\
2003 & Sep & 4128, 4153, 4157 & 4136, 4145 \\
2003 & Oct & 4184, 4188, 4198, 4207 & 4187, 4192, \underline{4203} \\
2003 & Nov & 4253 & 4247, 4263, 4288 \\
 & & & \\
2004 & Aug & 4797 & - \\
2004 & Sep & 4858 & - \\
2004 & Oct & 4868, 4874, 4895, 4905, 4917 & - \\
2004 & Nov & 4933, 4948 & 4930 \\
2004 & Dec & - & 5042, 5052 \\
 & & & \\
2005 & Sep & 5566, 5603, 5610, 5622, 5633, 5642, 5658 & 5582, 5597, 5607, 5619, 5628, 5637, 5646, 5666 \\
2005 & Oct & 5709, 5719, 5731, 5744, 5759, 5765, 5770, 5777, 5781, 5792, 5800 &
             5675, 5681, 5713, 5729, 5730, 5732, 5745, 5760, 5763, 5771, 5776, 5782, 5786, 5797 \\
2005 & Nov & 5813, \underline{5823}, 5842, 5865, 5866, 5872, 5878, 5898, 5902, 5918 &
             5807, 5820, 5836, 5847, 5853, 5870, 5871, 5882, 5889, 5895, 5905, 5924 \\
\hline
\end{tabular}}
\label{tab:runs}
\end{table*}

One meaning of the verb ``to vary'' is to change in amount or level, especially from one occasion to another.
In astronomy, there are many types of temporal variability. Stars may change in brightness, in which
case they are termed ``variables'', or they may change position on the sky, in which case they have a proper motion.
Over a sufficiently long period of time, the shape of constellations change.
Galaxies exhibit variability across the electromagnetic spectrum since their emission is made up of the radiation from billions of sources,
although their most obvious source of variation comes from active galactic nuclei or supernovae.
Transient events, such as gamma-ray bursts, or microlensing events, are instrinsically variable.
Solar system objects from planets to asteroids drift slowly across the sky, waxing and waning on various timescales.
In fact, at some level, everything in the sky is variable.

The introduction of CCD detectors to astronomy greatly enhanced the ability to conduct variability surveys.
The extension of CCD cameras to mosaic and wide-field formats along with the exponential progression of
computing power have allowed the subsequent development of more ambitious surveys reaching to
deeper magnitudes, higher cadences and larger sky areas. For more details we direct the
reader to \citet{bec2004} who present a clear summary of modern variability surveys.
In this work we concentrate on optical photometric and astrometric variability (hence ``light and motion'')
over a $\sim$249~deg$^{2}$ patch of sky.

Large sky surveys such as the Sloan Digital Sky Survey~(SDSS; \citealt{yor2000}) have in many
ways revolutionised our knowledge of the Universe. SDSS has imaged approximately a quarter of the sky
in five photometric wave bands. The exploitation of this impressive dataset has resulted in
hundreds of publications covering a wide range of astronomical topics, from the structure of the
Milky Way to the mapping of a large fraction of the Universe. The bulk of this data, however, contain only
single measurements of objects from the north Galactic cap with no information on possible photometric
variability or astrometric motion.
Substantial efforts have been made by \citet{mun2004} (see also \citet{gou2004}) to measure proper motions by matching
SDSS data from the north Galactic cap with the USNO-B catalogue (\citealt{mon2003}).
The resultant proper motion catalogue is 90\% complete down to $g~=~19.7$ with the
magnitude limit being set by the USNO-B catalogue faint magnitude limits.

One of the primary goals of the SDSS is the study of the variable
sky (\citealt{ade2007}) of which our knowledge is still very incomplete (\citealt{pac2000}).
To this end, the SDSS has repeatedly imaged a 300 square degree area, the so called
Stripe 82, during the later half of each year since 1998. In 2005, the SDSS-II Supernova Survey
(\citealt{fri2008}) started with the aim of detecting Type-I supernovae in Stripe 82, greatly improving
the cadence of measurements within the stripe.
By averaging a subset of the repeated observations of unresolved sources in Stripe 82, \citet{ive2007} built a
standard star catalogue containing $\sim$1 million nonvariable sources with
$r$ band magnitudes in the range 14-22, by far the deepest and most numerous set of photometric
standards available. Using these same multi-epoch photometric data, \citet{ses2007}
analysed the photometric variability for $\sim$1.4~million unresolved sources in the stripe, drawing
interesting conclusions on the spatial distribution of RR Lyrae stars and the variability of quasars.

Here we present a new public archive of light-motion curves in SDSS Stripe 82.
The archive has been constructed from the set of high-precision multi-epoch photometric and astrometric
measurements made in the stripe since the first SDSS runs in 1998 until the end of 2005.
In constructing the catalogue, we only use measurements of objects that are cleanly detected 
in individual SDSS runs.
The catalogue contains almost 4~million objects, galaxies and stars,
and is complete down to magnitude 21.5 in $u$, $g$, $r$ and $i$, and to magnitude 20.5 in $z$.
Each object has its proper motion calculated based only on the multi-epoch SDSS J2000 astrometric
measurements. The catalogue reaches almost two magnitudes deeper than the SDSS/USNO-B catalogue, making
it the deepest large-area photometric and astrometric catalogue available.

The catalogue comes in two flavours,
the Light-Motion Curve Catalogue (LMCC), which contains the set of individual
light-motion curves, where measured quantities for each object are listed as a function of
wave band and epoch, and the Higher-Level Catalogue (HLC), which presents
a set of derived quantities for each light-motion curve. For many purposes, it
is more convenient to work with the HLC, especially for selecting subsets of
interesting objects. The construction, calibration and format of the LMCC
are discussed in Section~2, and the HLC is described in Section~3.
Between the two subcatalogues, there is all the necessary
information available to explore the photometric and astrometric variability of
$\sim$249 square degrees of equatorial sky.
In Section~4, we investigate the quality of the photometric and astrometric properties of our
catalogues by comparing them against suitable external catalogues, and we analyse
the behaviour of our proper motion uncertainties.

\section{The Light-Motion Curve Catalogue}

\subsection{Stripe 82 Data}

The SDSS photometric camera is mounted on a 2.5m dedicated telescope at the Apache Point
Observatory, New Mexico. It consists of a photometric array of 30 SITe/Tektronix CCDs, each of size 2048x2048 pixels, arranged in the focal
plane of the telescope in six columns of five chips each (\citealt{gun1998}; \citealt{gun2006}) with a space of approximately 
one chip width between columns. Each row of six chips
is positioned behind a different filter so that SDSS imaging data is produced in five wave bands, namely, $u$, $g$, $r$, $i$ and $z$
(\citealt{fuk1996}; \citealt{smi2002}). The camera operates in time-delay-and-integrate (TDI) drift-scan mode at the sidereal rate and the chip
arrangement is such that two scans cover a filled stripe 2\fdg54 wide, with $\sim$1\arcmin$\,$ overlap between chip columns in the two scans.
In addition, the camera contains an array of 24 CCDs with 400x2048 pixels which enable observations of bright astrometric 
reference stars for subsequent astrometry and focus monitoring.

The images are automatically processed through specialised pipelines (\citealt{lup1999}; \citealt{lup2001};
\citealt{hog2001}; \citealt{sto2002}; \citealt{ive2004})
producing corrected images, object catalogues, astrometric solutions, calibrated fluxes and many other
data products. The object catalogues, which include the calibrated photometry and astrometry, are stored in FITS binary table format
(\citealt{wel1981}; \citealt{cot1995}; \citealt{han2001}) and referred
to as ``tsObj'' files. It is these object catalogues that we have used to construct the LMCC.

The SDSS Stripe 82 is defined as the region spanning 8 hours in right ascension (RA)
from $\alpha = 20^{\mbox{\small h}}$ to 4$^{\mbox{\small h}}$ and
2\fdg5 in declination (Dec) from $\delta = -$1\fdg25 to 1\fdg25.
The stripe consists of two scan regions referred to as the north and south strips. Both the north and
south strips have been repeatedly imaged from 1998 to 2005, between June and December of each year, with 62 of the 134 imaging runs obtained in 2005
alone (this large sampling rate was produced by the start of the SDSS-II Supernova Survey).
A specific imaging run may cover all of one strip or some fraction of the area,
and images of the same patch of sky are never taken more than once per night. Hence the exact temporal coverage and cadence of the
light-motion curves in the catalogue are strong functions of celestial position.
In Table~\ref{tab:runs}, we list the SDSS imaging runs included in the LMCC organised by the month and year of observation. Not all scans of Stripe 82 were
included in our analysis due to failures in the SDSS {\tt frames} pipeline (\citealt{lup2001})
when processing a run or failure of our calibration routines to produce photometric zeropoints (see Section~2.2).

\subsection{Further Photometric Calibrations}

The Stripe 82 data set includes 62 ``standard'' SDSS imaging runs
which were observed under photometric conditions and which
were photometrically calibrated using the standard SDSS pipelines (\citealt{tuc2006}).
We use these standard runs to construct a reference catalogue
of bright star fluxes, from which we can both improve the photometric
calibrations of the standard runs, as well as derive photometric
calibrations for the Stripe 82 supernova imaging runs from 2005, which were
generally observed under non-photometric conditions.

To construct the reference catalogue, we start with a set
of bright, unsaturated stars, with $14 < r < 18$, taken from
a set of high quality runs acquired over an interval of less than twelve months (2659, 2662, 2738, 2583, 3325, 3388).
We then match the individual detections of these stars in each of the
62 standard runs, using a matching radius of 1 arcsec.  On average,
there are 10 independent measurements of each star among the standard runs,
and we only include in the reference catalogue stars with 5 or more
measurements.  We then compute the unweighted mean of the independent flux
measurements of each star and adopt that mean flux in our
reference catalogue. Note that we specifically use the fluxes measured in the
so-called SDSS ``aperture 7'', which has a radius of 7.43 arcsec;
this aperture is typically adopted in the SDSS as a reference aperture
appropriate for isolated bright star photometry.

\begin{figure}
\epsfig{file=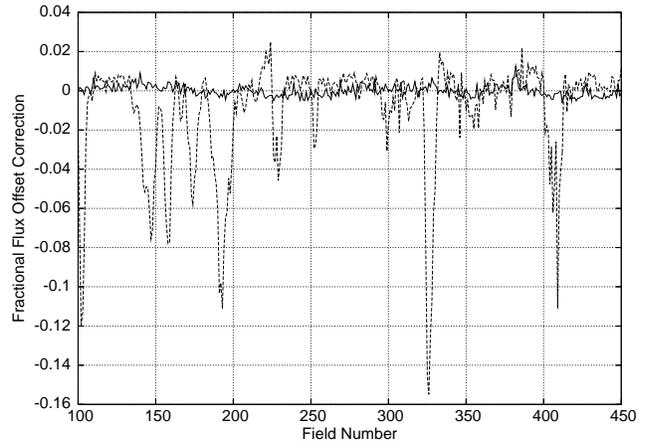,angle=270.0,width=\linewidth}
\caption{Plot of the median fractional flux offset of the reference stars relative to the reference catalogue as a function of
         field number for runs 94 (solid line) and 5853 (dashed line). For both runs we show the offsets for camera column 1 and the $r$ band.
         Run 94 is a typical pre-2005 run observed under
         photometric conditions and run 5853 is a typical 2005 supernova run observed under non-photometric conditions. \label{fig:zp}}
\end{figure}

\begin{table*}
\centering
\caption{The list of extra constraints that all need to be satisfied in at least one wave band in order for an object record to be included in the LMCC.}
\begin{tabular}{@{}lcccll}
\hline
Tag Name In & Relation & Flag              & Value & Flag Name & Description \\
tsObj File  &          & (Hexadecimal Bit) &       &           &             \\
\hline
{\tt OBJC\_FLAGS} & AND & 0x4        & FALSE & {\tt EDGE}               & Reject objects too close to the edge of the image \\
{\tt FLAGS}       & AND & 0x10000000 & TRUE  & {\tt BINNED1}            & Accept only objects detected in the unbinned image \\
{\tt FLAGS}       & AND & 0x20       & FALSE & {\tt PEAKCENTER}         & Reject objects where the given centre is the position of the peak \\
                  &     &            &       &                          & pixel, rather than based on the maximum likelihood estimator \\
{\tt FLAGS}       & AND & 0x80       & FALSE & {\tt NOPROFILE}          & Reject objects that are too small or too close to the edge to \\
                  &     &            &       &                          & estimate a radial profile \\
{\tt FLAGS}       & AND & 0x40000    & FALSE & {\tt SATUR}              & Reject objects with one or more saturated pixels \\
{\tt FLAGS}       & AND & 0x80000    & FALSE & {\tt NOTCHECKED}         & Reject objects with pixels that were not checked to see whether \\
                  &     &            &       &                          & they included a local peak, such as the cores of saturated stars \\
{\tt FLAGS}       & AND & 0x400000   & FALSE & {\tt BADSKY}             & Reject objects with a sky level so badly determined that the highest \\
                  &     &            &       &                          & pixel in an object is {\it very} negative, far more so than a mere \\
                  &     &            &       &                          & non-detection \\
{\tt FLAGS2}      & AND & 0x100      & FALSE & {\tt BAD\_COUNTS\_ERROR} & Reject objects containing interpolated pixels that have too few \\
                  &     &            &       &                          & good pixels to form a reliable estimate of the flux error \\
{\tt FLAGS2}      & AND & 0x800      & FALSE & {\tt SATUR\_CENTER}      & Reject objects with a centre close to at least one saturated pixel \\
{\tt FLAGS2}      & AND & 0x1000     & FALSE & {\tt INTERP\_CENTER}     & Reject objects with a centre close to at least one interpolated pixel \\
{\tt FLAGS2}      & AND & 0x4000     & FALSE & {\tt DEBLEND\_NOPEAK}    & Reject child objects with no detected peak \\
{\tt FLAGS2}      & AND & 0x8000     & FALSE & {\tt PSF\_FLUX\_INTERP}  & Reject objects with more than 20\% of the PSF flux from \\
                  &     &            &       &                          & interpolated pixels \\
\hline
\end{tabular}
\label{tab:req}
\end{table*}

These reference catalogue stars are used to re-calibrate the
standard runs, as well as to calibrate the 2005 supernova data.
For the supernova runs, we first adopt a sensible but arbitrary zeropoint
for the purposes of generating the initial tsObj files using standard
SDSS pipeline tools.
We then match the object detections in each run to the
reference catalogue, and compute the median fractional flux
offset of the reference stars in the individual run relative to the reference
catalogue\footnote{The $u$ band images possess a significantly
poorer signal-to-noise ratio than the other bands and the use of reference catalogue
stars that have a magnitude fainter than 18 {\it before} calibration (due to higher than usual extinction) degrades the
determination of the derived flux offsets. The flux offsets
for the $u$ band have therefore been determined using only the reference catalogue stars with
an uncalibrated magnitude brighter than 18 in each run.}
These median offsets are computed in two iterations.
We first calculate the median fractional offset for each run in bins of
0.0208$^{\circ}$ in Dec, i.e., 120 bins over the width of Stripe 
82, or about 10 bins per CCD width.  This exercise is designed to correct
flatfielding errors for a given run.  Note that these errors would only
depend on Dec because the SDSS employs a drift-scan camera,
and the scan direction for Stripe 82 is in the RA direction.
After correcting for the declination-dependent offsets,
we then re-compute the median fractional flux offsets for each
field along a given run (each SDSS field is 0.15$^{\circ}$ long in RA).
This additional field-by-field offset corrects for
any temporal variations in the photometric zeropoint of a given run,
which are due to transparency/extinction changes
over the course of a nominally photometric night.

Flux offsets for a certain wave band were only applied to objects assigned to a bin with at least 9 reference catalogue stars in order to guarantee
the accuracy of the derived flux offset. In practice, this extra restriction only affects the photometry of objects in the $u$
band, and in other wave bands when the atmospheric transparency is low. We use a photometric calibration tag
(see Table~\ref{tab:lmc}) to monitor whether or not a flux offset has been applied to the
photometry of a particular object at a certain epoch in a specific wave band.
The final tsObj files used for subsequent analyses therefore have
both these declination-dependent and field-dependent flux offsets removed for most object records.
We find that for the standard SDSS runs, the fractional flux
offset corrections, which we refer to as photometric zeropoints, are about $1-2\%$, which sets the typical scale of these
residual errors in the standard SDSS calibration procedures. Figure~\ref{fig:zp} shows the fractional flux offset corrections
as a function of SDSS field number (an arbitrary coordinate along RA assigned to image sections from the same run)
for a typical photometric run (94) and a typical non-photometric run (5853).

\subsection{Catalogue Construction}

The object catalogues (tsObj files) contain quality and type flags for each object record to 
aid in the selection of ``good'' measurements and specific data samples. In the LMCC, we only accept
object records classified as galaxies/non-PSF-like objects (tsObj file tag {\tt OBJC\_TYPE} $=$ 3) or stars/PSF-like objects 
(tsObj file tag {\tt OBJC\_TYPE} $=$ 6), and
the object must have no child objects (tsObj file tag {\tt NCHILD} $=$ 0; \citet{sto2002}). We then require that an object
record satisfies all of a set of constraints in at least one wave band. The first of these constraints is that
a photometric zeropoint, calculated using the method described in Section~2.2, has been applied to the object record,
and that the object record has an {\it uncalibrated} PSF magnitude 
(tsObj file tag {\tt PSFCOUNTS}) brighter than 21.5 for the bands
$u$, $g$, $r$ and $i$, or brighter than 20.5 for the $z$ band. These limits were chosen to ensure
that any photometric measurement in the LMCC has a signal-to-noise ratio of at least 5 in at least one wave band.
In Table~\ref{tab:req}, we list the remaining set of constraints to be satisfied in at least one wave band in order for an object record to be 
included in the LMCC. 

We apply one final constraint on the quality of an object record in order to avoid the inclusion of 
cosmic ray events in our catalogue. If an object record satisfies all of the above constraints
in one wave band only, then it is accepted only if the tsObj file tag {\tt FLAGS2} for that wave band does not contain the hexadecimal
bit 0x1000000 (flag name {\tt MAYBE\_CR}), the presence of which indicates that the object is possibly a cosmic ray.

\begin{figure*}
\def\subfigtopskip{4pt}   
\def\subfigbottomskip{8pt}
\def\subfigcapskip{4pt}
\centering
\epsfig{file=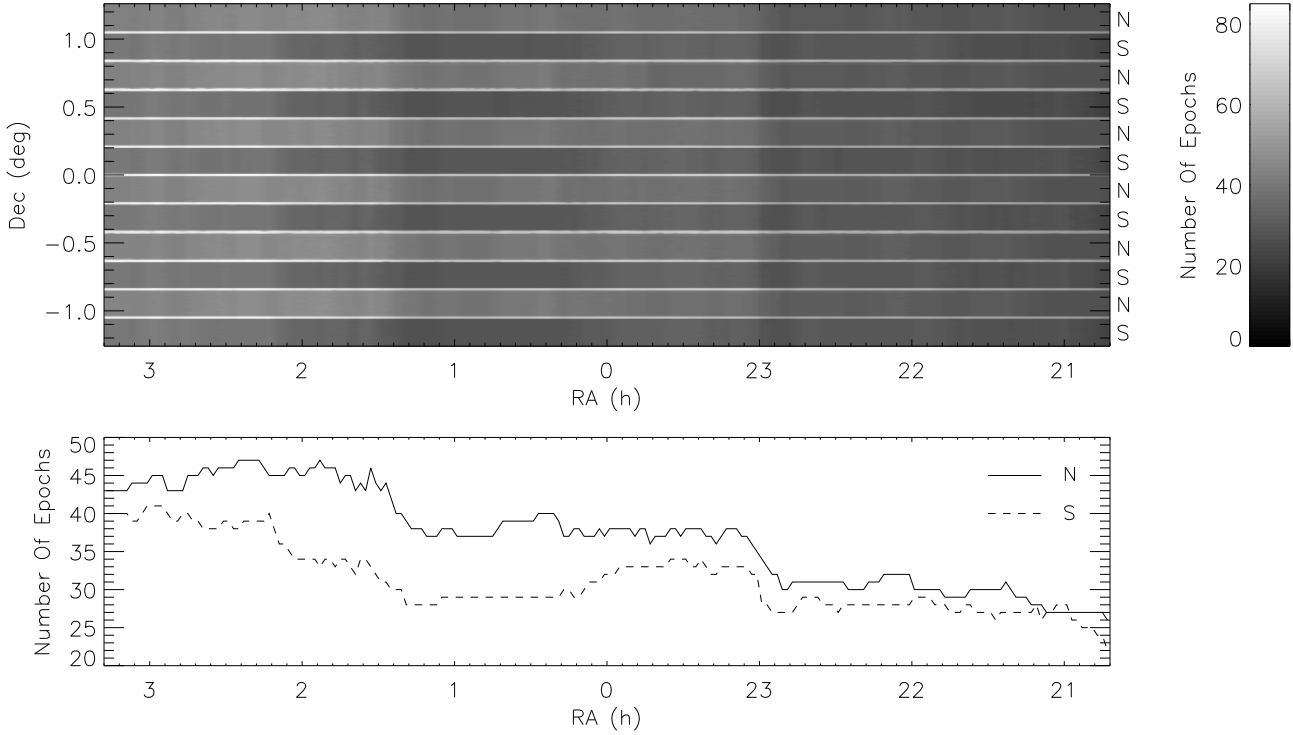,angle=0.0,width=\linewidth}
\caption{{\bf Top panel:} Greyscale image showing the maximum number of epochs in the light-motion curves as a function of object position.
         {\bf Bottom panel:} Maximum number of epochs in the light-motion curves as a function of RA for a 0\fdg01 wide slice through the
         greyscale image centred at $\delta~=~-0\fdg1$ (continuous line; slice through the North strip) and for a similar slice centred
         at $\delta~=~0\fdg1$ (dashed line; slice through the South strip).
         \label{fig:tcov}}
\end{figure*}

\begin{figure*}
\def\subfigtopskip{4pt}
\def\subfigbottomskip{8pt}
\def\subfigcapskip{4pt}
\centering
\epsfig{file=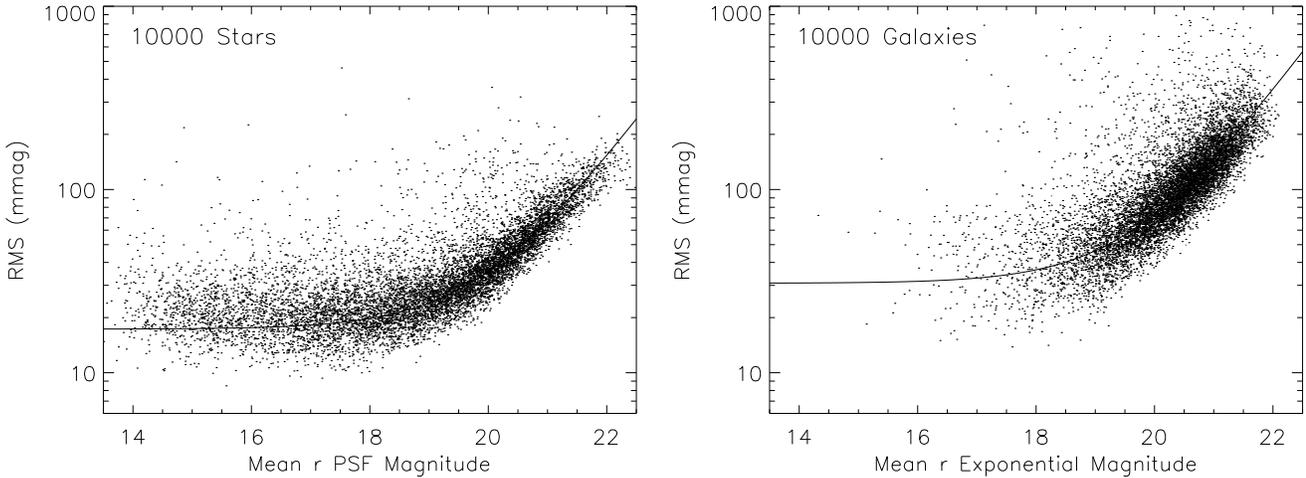,angle=0.0,width=\linewidth}
\caption{{\bf Left:} Plot of RMS PSF magnitude deviation versus mean PSF magnitude for
         10000 random PSF-like objects (stars) with at least 20 good $r$ magnitude measurements.
         {\bf Right:} Plot of RMS exponential magnitude deviation versus mean exponential magnitude for
         10000 random non-PSF-like objects (galaxies) with at least 20 good $r$ magnitude measurements.
         {\bf Both panels:} Plotted functions (continuous lines) are all of the form $f(m)~=~A~+~B~\,\exp~(C~\,~(m~-~18))$ where $m$
         denotes magnitude and $A$, $B$ and $C$ are fitted parameters. For the left hand panel, $A$, $B$ and $C$ have values
         17.3~mmag, 2.48~mmag and 1.00, respectively. For the right hand panel, $A$, $B$ and $C$ have values
         30.8~mmag, 5.67~mmag and 1.01, respectively.
         \label{fig:rms_phot}}
\end{figure*}

In order to construct the light-motion curves, we processed each run in turn, starting with the 2005 runs which were
closely spaced in time. For each object record in the current run satisfying our quality and type constraints,
we used the following algorithm to process the record:
\begin{enumerate}
  \item We define a subset of objects with light-motion curves from the current catalogue that have mean positions
        inside a 1\arcmin$\,$ box centred on the position of the current object record.
  \item We calculate an expected position at the epoch of the current object record for each object in the subset
        using one of two methods depending on the number of epochs in the corresponding light-motion curve.
        If an object has a light-motion curve with six epochs or less, then the mean position is used for the
        expected position. Otherwise a mean position and proper motion are fitted to the light-motion curve
        and used to calculate the expected position of the object at the epoch of the current object record.
  \item From the expected positions of the subset of objects, we find the closest object to the position of the current object
        record\footnote{Note that an object record contains a single datum for the celestial coordinates, calculated from the
        astrometric solution for the SDSS CCD camera at the current epoch.}.
        If the closest object lies within 0.7\arcsec, then the current object record is appended to the
        light-motion curve of the closest object, otherwise a new light-motion curve is created containing only the
        current object record.
\end{enumerate}
Since each run contains at most one measurement of any object, the above algorithm can be performed in parallel for all objects from one run.

We choose to include both the PSF magnitude and exponential-profile magnitude (\citealt{sto2002}) in a light-motion curve
as a measure of an object's brightness in each wave band at each epoch. The PSF magnitude is the optimal measure
of the brightness of a point-source object, and hence it is suitable for studying stars and quasars.
Photometry of extended objects, such as galaxies, may be performed in a variety of ways, including fitting an exponential profile
to the object image. The advantage of including the exponential magnitude as opposed to any of the other available profile magnitudes
is that the difference between the PSF and exponential magnitudes, referred to as a concentration index, may be used as a continuous
object-type classifier (\citealt{scr2002}), independent of the more restrictive binary SDSS classification.

After processing all Stripe 82 data, the LMCC was trimmed to only those objects that have a mean position in the range 
$\alpha = 20.7^{\mbox{\small h}}$ to 3.3$^{\mbox{\small h}}$ and $\delta = -$1\fdg26 to 1\fdg26.
This was desirable because the temporal coverage is too sparse outside these limits.
The total area of sky covered by the LMCC is therefore $\sim$249 deg$^{2}$.
The LMCC was also searched for photometric outliers using a 3$\sigma$-clip algorithm for outlier identification, 
and we found clear groups of outliers clustered at specific HJDs. The tight clustering in time indicates that these groups of outliers are
simply due to bad epochs (and not due to some astrophysical process, like an eclipse or flare), and hence we removed
the corresponding data points, which amounted to $\sim$1\% of all epochs. 

\begin{figure}
\def\subfigtopskip{4pt}
\def\subfigbottomskip{8pt}
\def\subfigcapskip{4pt}
\centering
\begin{tabular}{c}
\subfigure[]{\epsfig{file=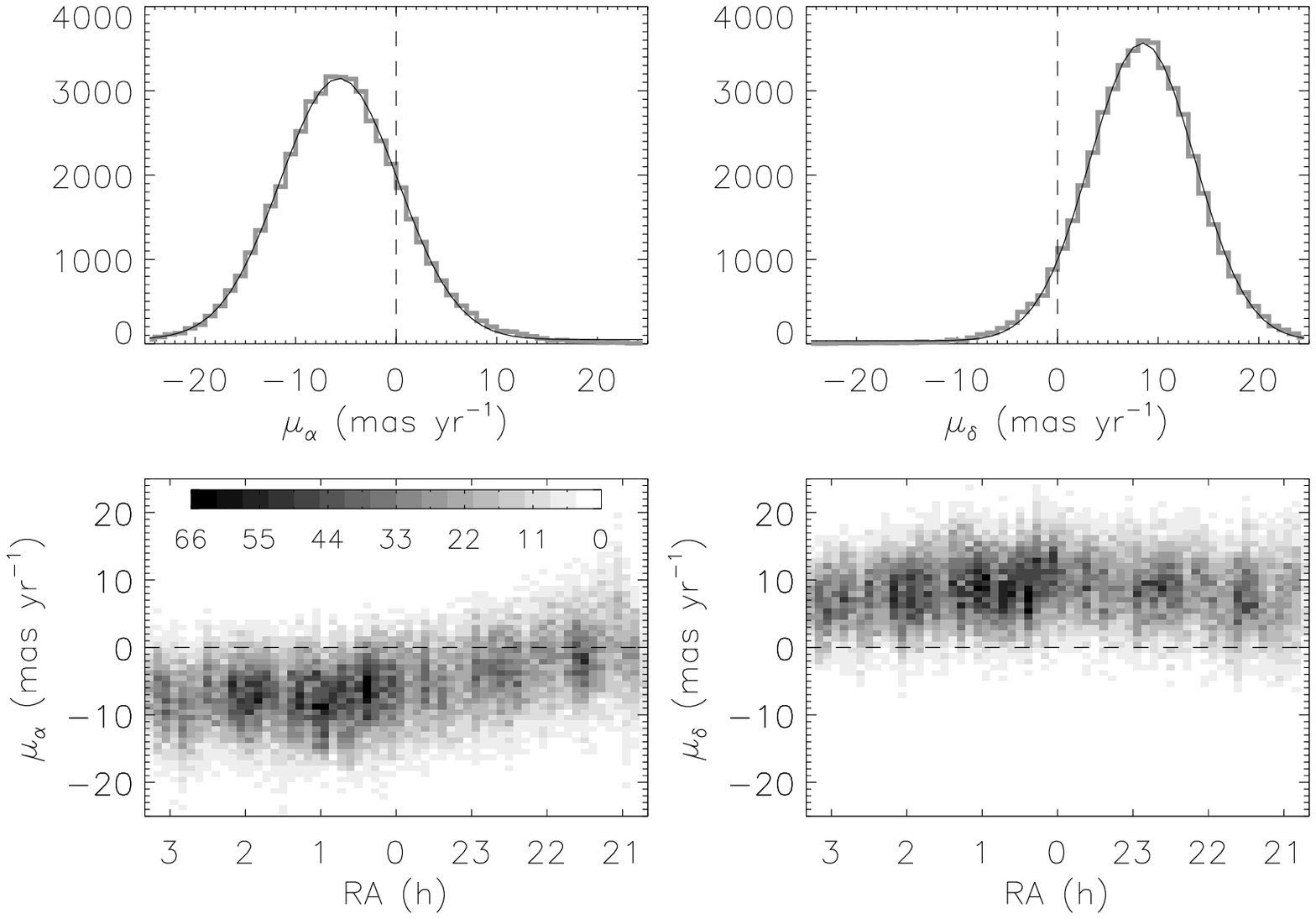,angle=0.0,width=\linewidth} \label{fig:astro_sysa}} \\
\subfigure[]{\epsfig{file=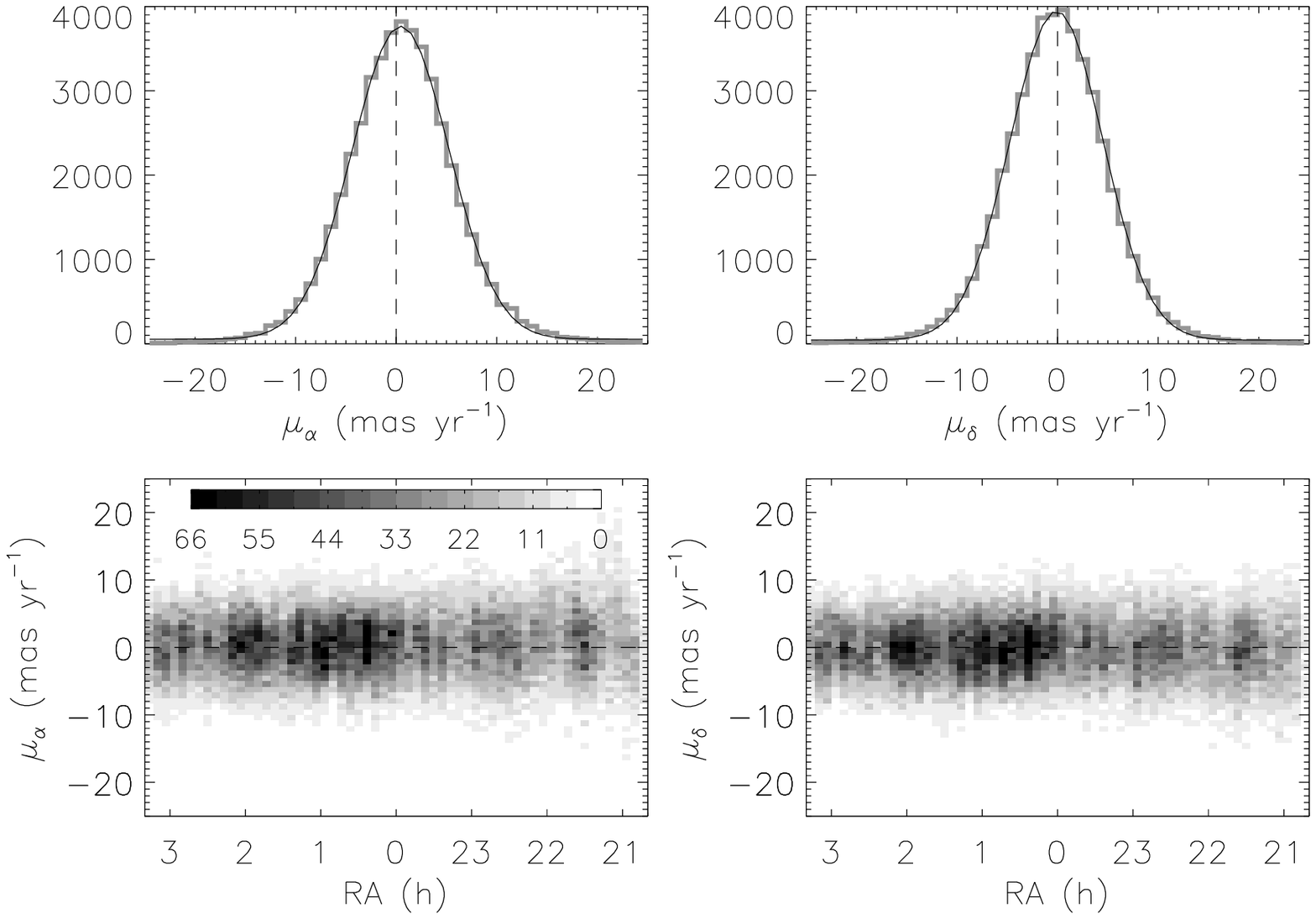,angle=0.0,width=\linewidth} \label{fig:astro_sysb}} \\
\end{tabular}
\caption{{\bf (a): Top row:} Histograms of galaxy proper motions in RA (left) and Dec (right) {\it before}
              astrometric recalibration.
              {\bf Bottom row:} Density plots of galaxy proper motions in RA (left) and Dec (right)
              as a function of RA {\it before} astrometric recalibration. The intensity bar has units of
              number of galaxies deg$^{-1}$ mas$^{-1}$ yr.
         {\bf (b):} The same as Figure~\ref{fig:astro_sysa} but {\it after} astrometric recalibration.
         \label{fig:astro_sys}}
\end{figure}

\begin{figure}
\epsfig{file=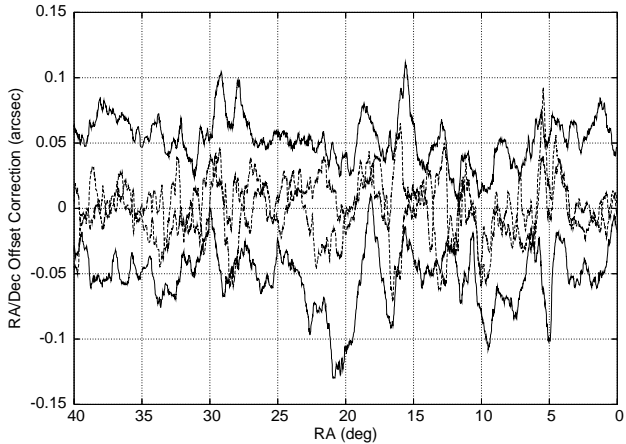,angle=270.0,width=\linewidth}
\caption{Plot of the mean offsets in RA and Dec relative to the reference run 5823
         as a function of RA for runs 94 (solid lines) and 5918 (dashed lines). For both runs
         we show the offsets for camera column 1, and for run 94, the lower and upper lines correspond
         to the RA and Dec offsets, respectively. Note that this is but a small section of the full stripe.
         \label{fig:ac}}
\end{figure}

In the top panel of Figure~\ref{fig:tcov}, we plot a greyscale image showing the maximum number of epochs in the light-motion curves
as a function of object position. One may clearly see that the light-motion curves for objects in the overlap regions 
between the north and south strips contain approximately twice as many epochs as the light-motion curves for objects elsewhere.
In the bottom panel, we plot a one-dimensional slice through the greyscale image for $\delta~=~-0\fdg1$ and $\delta~=~0\fdg1$ to 
further illustrate the dependence of the number of light-motion curve epochs on RA.

In Figure~\ref{fig:rms_phot}, we plot sample photometric RMS diagrams for the $r$ band. 
The left hand panel shows the RMS PSF magnitude deviation versus mean PSF magnitude for 10000 random PSF-like objects
({\tt MEAN\_OBJECT\_TYPE~=~6}; see Section~3.1) 
that have light-motion curves with at least 20 good $r$ magnitude measurements. PSF-like objects are mainly stars with some 
contamination by quasars.
Similarly, the right hand panel shows the RMS exponential magnitude deviation versus mean 
exponential magnitude for 10000 random non-PSF-like objects ({\tt MEAN\_OBJECT\_TYPE~=~3}) that have light-motion curves with at least
20 good $r$ magnitude measurements. Non-PSF-like objects are mainly galaxies. 

Overplotted on each RMS diagram is an empirical fit to the data using an exponential function
of the form $f(m)~=~A~+~B~\,\exp~(C~\,~(m~-~18))$ where $m$ denotes magnitude and $A$, $B$ and $C$ are fitted parameters
(whose values are reported in the caption of Figure~\ref{fig:rms_phot}). 
The fitting was done using an iterative 3$\sigma$-clip algorithm
(see Vidrih index under Section~3.1). It is clear from these diagrams that for the $r$ band we are achieving 
$\sim$20~mmag and $\sim$30~mmag RMS accuracy at $r\sim$18~mag for PSF-like and non-PSF-like objects, 
respectively\footnote{The upturn in the RMS amplitude for stars brighter than $r~\sim$~16 is due to a
combination of factors including: the appearance of detectable asymmetric
low surface-brightness structure (e.g. diffraction spikes) in the images;
the effect of the large angular size of the images on the determination
of the sky-background.}. 
The RMS diagrams for the other wave bands are very similar except for the $u$ band, where we achieve
$\sim$23~mmag and $\sim$50~mmag RMS accuracy at $r\sim$18~mag for PSF-like and non-PSF-like objects, respectively.

\subsection{Further Astrometric Calibrations}

Each SDSS imaging run was astrometrically calibrated against
the US Naval Observatory CCD Astrograph Catalog
(UCAC; \citealt{zac2000}), yielding absolute positions accurate to
$\sim$45~mas RMS per coordinate (\citealt{pie2003}).  The accuracy
is limited primarily by the accuracy of the UCAC
positions ($\sim$70~mas RMS at the UCAC survey limit of $R \simeq 16$),
as well as the density of UCAC sources.  The version of UCAC used to
calibrate SDSS lacked proper motions, thus any proper motions based on
SDSS positions will be systematically in error by the mean proper motion of
the UCAC calibrators.

We illustrate the systematic errors inherent in the SDSS astrometry by
considering galaxies in the magnitude range $17~<~r~<~19.5$ that have
light-motion curves with at least 20 astrometric measurements.
For this set of galaxies we measure the proper motion in RA
and Dec, and we find that the galaxies are systematically moving with proper motions
of the order of 10~mas per year in both right ascension and declination!
The problem is clearly evident in Figure~\ref{fig:astro_sysa}
where we show histograms of the galaxy proper motions in RA (top left hand panel)
and Dec (top right hand panel). The trends of galaxy proper motion with RA
are shown in the bottom left hand panel for proper motion in RA and in
the bottom right hand panel for proper motion in Dec.

\begin{figure*}
\def\subfigtopskip{4pt}
\def\subfigbottomskip{8pt}
\def\subfigcapskip{4pt}
\centering
\begin{tabular}{cc}
\subfigure[]{\epsfig{file=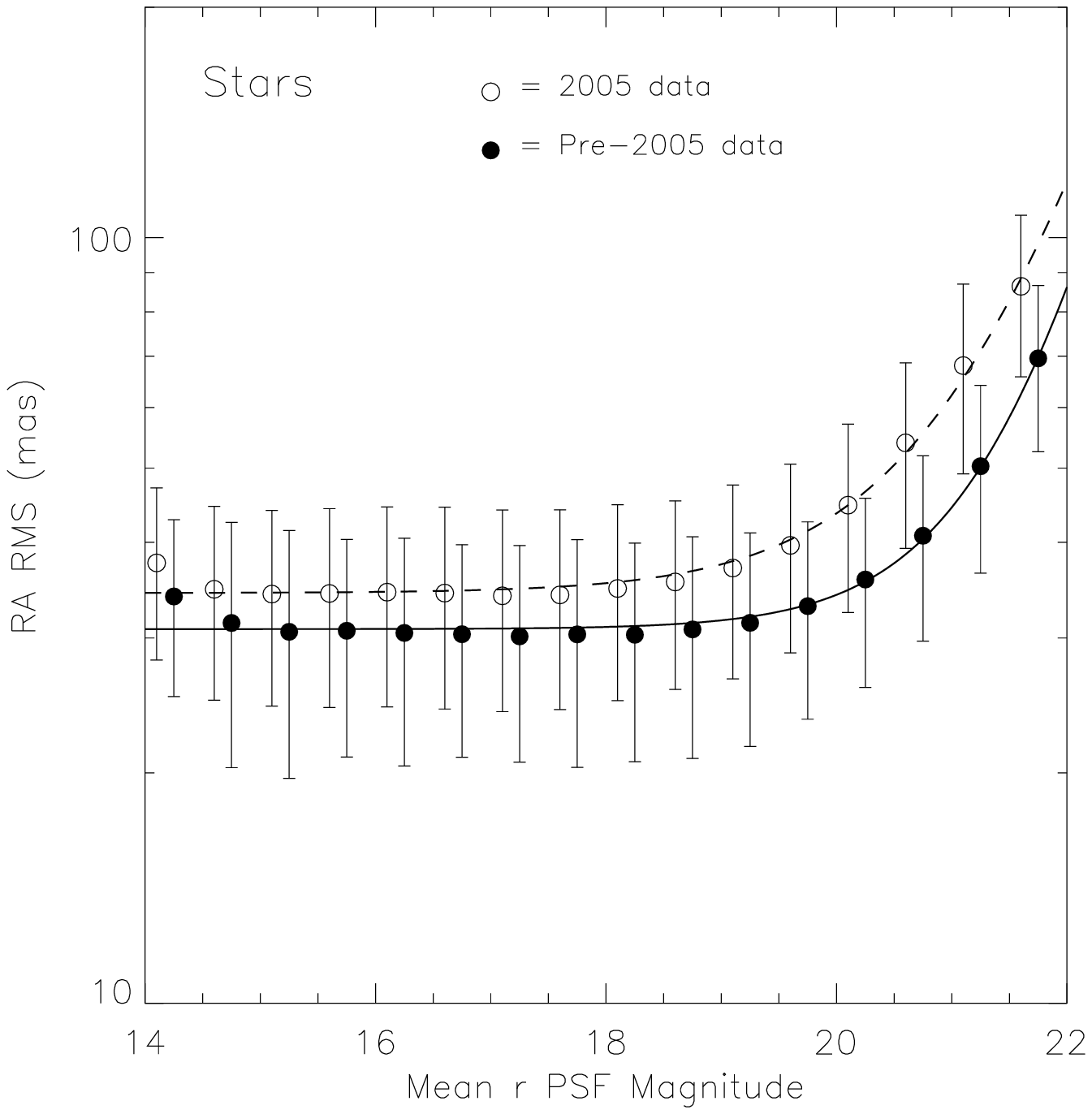,angle=0.0,width=0.5\linewidth} \label{fig:rms_astro_a}} &
\subfigure[]{\epsfig{file=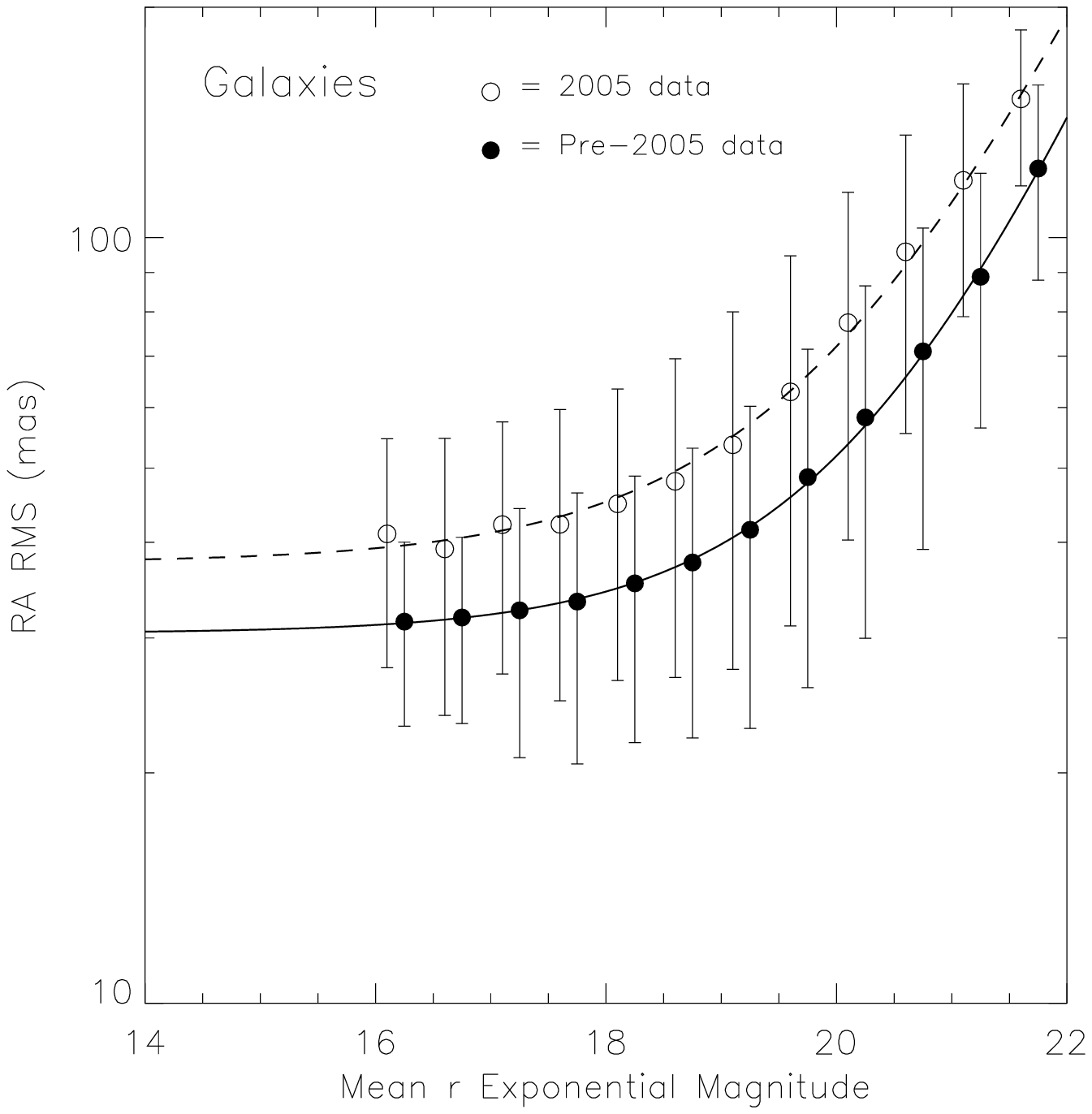,angle=0.0,width=0.5\linewidth} \label{fig:rms_astro_b}} \\
\end{tabular}
\caption{{\bf (a):} Plot of the peak RA RMS deviation in 0.5~mag bins versus $r$ magnitude for PSF-like objects (stars) for pre-2005
         data (filled circles) and 2005 data (open circles; offset by 0.15~mag to the left for clarity).
         {\bf (b):} Plot of the peak RA RMS deviation in 0.5~mag bins versus $r$ magnitude for non-PSF-like objects (galaxies) for pre-2005
         data (filled circles) and 2005 data (open circles; offset by 0.15~mag to the left for clarity).
         {\bf Both panels:} Error bars represent the dispersion in the distribution of RMS deviations in each magnitude bin.
         Plotted functions (continuous and dashed curves) are all of the form $f(m)~=~A~+~B~\,\exp~(C~\,~(m~-~18))$ where $m$
         denotes magnitude and $A$, $B$ and $C$ are fitted parameters.
         \label{fig:rms_astro}}
\end{figure*}

Proper motions based on multi-epoch SDSS data can be improved by recalibrating
each SDSS imaging run against a reference SDSS run, rather than using the
UCAC catalogue positions. The accuracy of the relative astrometry between runs is
$\sim$20 mas RMS per coordinate (\citealt{pie2003}), far superior to
the accuracy of the absolute positions, due both to the more accurate
centroids, as well as the far greater density of calibrators, for SDSS compared
to UCAC. Further, by using galaxies as calibrators, the proper motions can be
tied to an extragalactic reference frame and are thus inertial. This is
the method used in this paper to correct the systematic SDSS astrometric
errors illustrated in Figure~\ref{fig:astro_sysa}. However, it is
appropriate to mention that the positional system of the calibrators
still refers to the epochs given by UCAC.

All imaging runs along the north strip have been recalibrated using run
5823 as the reference run.  All imaging runs along the south strip have
been recalibrated against run 4203, after first recalibrating run 4203 against
run 5823. In order to recalibrate a target run, offsets in RA and
Dec are calculated for matching ``clean'' galaxies in the reference run 
(rejecting galaxies affected by problems with deblending, pixel interpolation, multiple
matches, etc. in either run). Only galaxies in the magnitude range $17~<~r~<~19.5$ are used
to avoid large galaxies with poorly defined centroids.
For each object in the target run, the mean offsets in 
RA and Dec for the nearest\footnote{Nearest in coordinate parallel to the scan direction, right ascension -- the coordinate
perpendicular to the scan direction, declination, is ignored, as the length of the binning
window is always larger than the width of a scan.}
100 such galaxies are calculated and added to
the object position. This procedure recalibrates the positions in the target run to the reference
frame defined by the galaxies in the reference run.

In Figure~\ref{fig:ac}, we show example mean offsets in RA and Dec for camera column 1 from runs 94 and 5918.
Notice that run 94, observed in 1998, requires larger mean offsets to correct for the mean proper motion of the UCAC calibration stars
than run 5918, observed in 2005, since it is further away in time from when the reference run 5823 was observed in 2005.

After recalibrating the astrometry for all light-motion curves in the LMCC, we have recreated Figure~\ref{fig:astro_sysa}
as Figure~\ref{fig:astro_sysb} using the same sample of galaxies. The histograms of the galaxy proper motions
in RA and Dec are now centred around $\sim$0~mas~yr$^{-1}$ indicating that galaxies are stationary in the recalibrated astrometric
system of the LMCC. Also, the lower panels demonstrate that the RA dependence of the galaxy proper motions has
been properly removed.

There is also some evidence that the galaxy proper motion scatter has been improved. The
RMS deviation of the residuals about a fourth-degree polynomial fit in each of the bottom panels of
Figure~\ref{fig:astro_sysa} is 5.4~mas~yr$^{-1}$ for RA and 5.2~mas~yr$^{-1}$ for Dec. This may be compared to the
improved RMS deviation of the proper motions in each of the bottom panels of Figure~\ref{fig:astro_sysb} at
4.8~mas~yr$^{-1}$ for RA and 4.6~mas~yr$^{-1}$ for Dec.

\begin{figure}
\def\subfigtopskip{4pt}
\def\subfigbottomskip{8pt}
\def\subfigcapskip{4pt}
\centering
\begin{tabular}{c}
\subfigure[]{ \epsfig{file=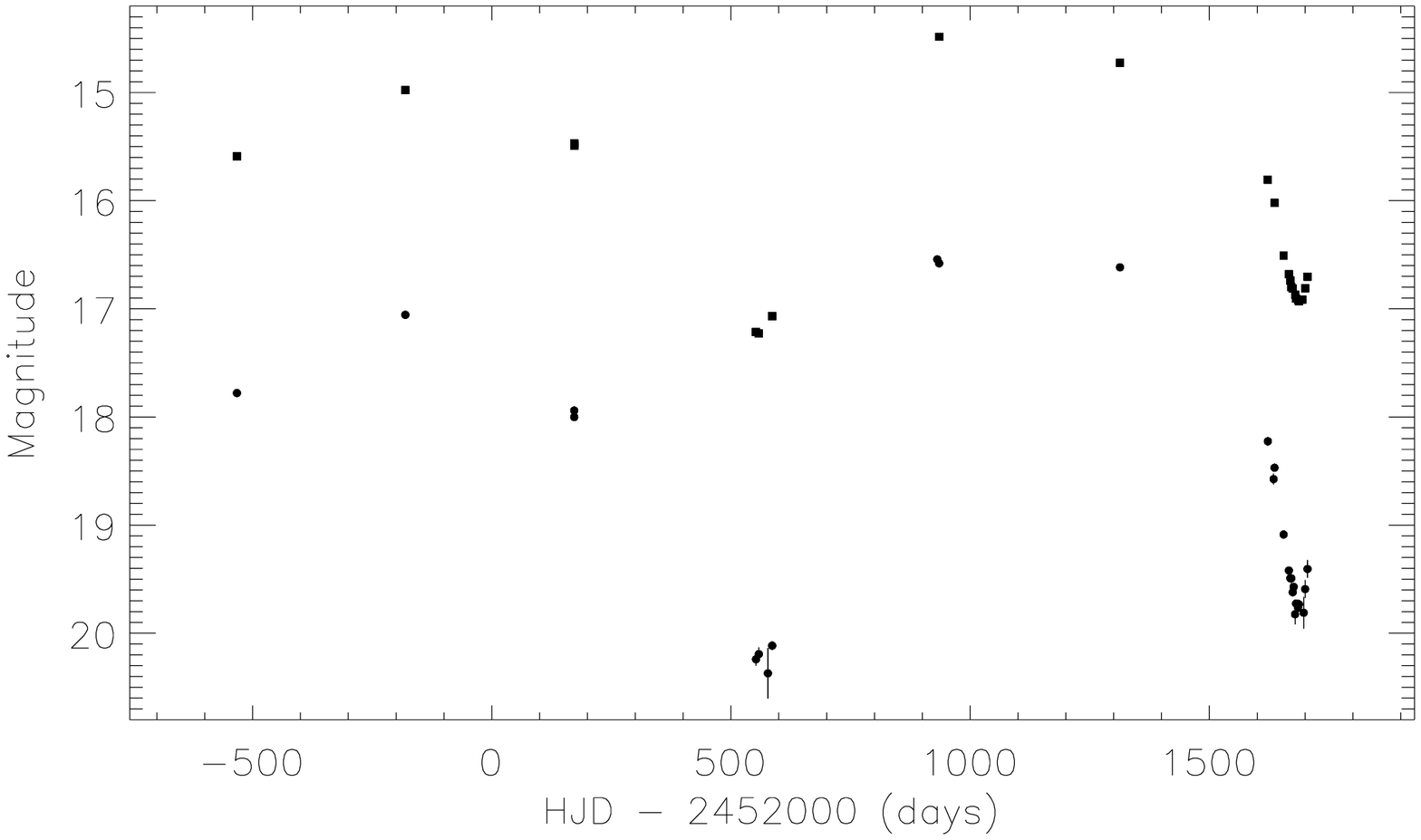,angle=0.0,width=\linewidth} \label{fig:lc}} \\
\subfigure[]{ \epsfig{file=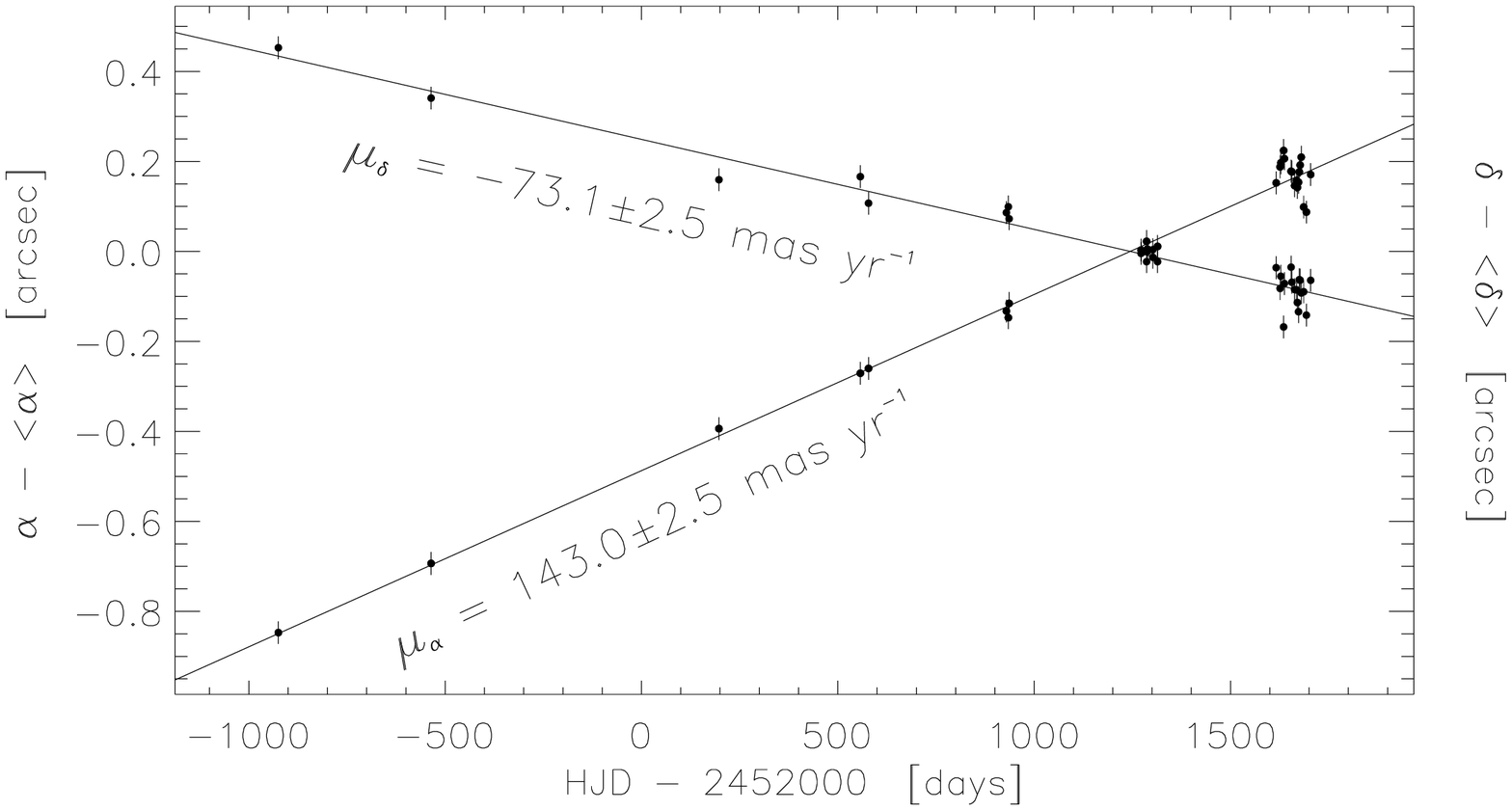,angle=0.0,width=\linewidth} \label{fig:pm}} \\
\end{tabular}
\caption{(a): Light curve of the Mira variable star candidate SDSS~J220514.58+000845.7 in the $r$ (upper points) and $g$ (lower points) bands.
         (b): Fitted proper motion in RA (track at bottom left, left-hand y-axis)
              and Dec (track at top left, right-hand y-axis) for the ultracool white dwarf
              SDSS~J224206.19+004822.7 (\citealt{kil2006}).
         \label{fig:example}}
\end{figure}

The SDSS pipelines do not supply uncertainties on the measured celestial coordinates in the tsObj files, and so we have determined a
noise model describing how the astrometric noise behaves as a function of magnitude. This was done by examining the distribution of
coordinate RMS for objects in the LMCC. However, we found that the astrometric noise in the 2005 observing season was noticeably
larger than in previous seasons, most likely due to the less stringent restrictions on observing conditions leading to a greater 
spread in PSF full-width half-maximum and object signal-to-noise. To properly account for this, we determined separate noise models
for the pre-2005 and 2005 observing seasons.

To determine the astrometric noise models, we select all PSF-like objects ({\tt MEAN\_OBJECT\_TYPE = 6}) with at least 20
good epochs in $r$ in each of the pre-2005 and 2005 observing seasons. For these objects we derive the distribution of
coordinate RMS deviations for 0.5~mag bins for both pre-2005 and 2005 data, and fit a peak and dispersion for each bin. 
In Figure~\ref{fig:rms_astro_a}, we plot the peak RA RMS deviation for each magnitude bin versus $r$ magnitude for pre-2005
data (filled circles) and 2005 data (open circles; offset by 0.15~mag to the left for clarity). 
We obtain very similar results for the Dec coordinate.
We fit the peak data as a function of magnitude $m$ via an exponential function of the
form $f(m)~=~A~+~B~\,\exp~(C~\,~(m~-~18))$ where $A$, $B$ and $C$ are fitted parameters, and plot the fitted models in 
Figure~\ref{fig:rms_astro_a} as continuous and dashed curves for pre-2005 and 2005 data, respectively. 

The following equations represent our final adopted astrometric noise model, based on the exponential model fits
for both the RA and Dec coordinates:
\begin{align}
\text{Pre-2005:} \,\, & \sigma_{\alpha}(t) = \sigma_{\delta}(t) = 32.0 + 0.430 \exp \left( 1.34 \left( m(t) - 18 \right) \right) \notag \\
\text{2005:}     \,\, & \sigma_{\alpha}(t) = \sigma_{\delta}(t) = 35.4 + 0.783 \exp \left( 1.09 \left( m(t) - 18 \right) \right) \notag \\
\label{eqn:sig_coord}
\end{align}
where $\sigma_{\alpha}(t)$ and $\sigma_{\delta}(t)$ are the uncertainties on the measured celestial coordinates
$\alpha(t)$ and $\delta(t)$, respectively, at time $t$, and
$m(t)$ represents the brightest PSF magnitude out of the five photometric measurements at time $t$.
Evidence that this noise model is valid comes from the fact that the distribution of $\chi^{2}$ per degree of 
freedom of the proper motion fit for the HLC (Section~3.1) is peaked at a value of $\sim$1.1.
Note that astrometric uncertainties are not given in the LMCC and should be obtained via Equation~\ref{eqn:sig_coord}.

In Figure~\ref{fig:rms_astro_b}, we plot the results of the same coordinate RMS deviation analysis for non-PSF-like objects
({\tt MEAN\_OBJECT\_TYPE~=~3}) with at least 20 good epochs in $r$ in each of the pre-2005 and 2005 observing seasons.
It is clear from the plots in Figure~\ref{fig:rms_astro} that we are achieving
$\sim$32~mas and $\sim$35~mas RMS accuracy at $r\sim$18~mag for stars for pre-2005 and 2005 data, respectively, and
$\sim$35~mas and $\sim$46~mas RMS accuracy at $r\sim$18~mag for galaxies for pre-2005 and 2005 data, respectively. 

\subsection{Catalogue Format}

The LMCC exists as eight {\tt tar} files, one for each hour in RA from 20$^{\mbox{\small h}}$ to
4$^{\mbox{\small h}}$. Each {\tt tar} file contains 60 subdirectories corresponding to the minutes of RA, and the light-motion curves
are stored in these directories based on their mean RA coordinates. The LMCC contains 3700548 light-motion curves, 
2807047 of which have at least
20 epochs. The {\tt tar} files ($\sim$29.5~Gb compressed) may be obtained by web
download from {\it http://das.sdss.org/value\_added/stripe\_82\_variability/SDSS\_82\_public/}. 
Light-motion curve plotting tools written in IDL may also be downloaded from the same website.

\begin{table*}
\centering
\caption{The list of columns that make up a light-motion curve in the LMCC along with a brief description.} 
\begin{tabular}{@{}llll}
\hline
Column Number & Column Name & Type & Description \\
\hline
1  & Run & INTEGER & SDSS imaging run \\
2  & Rerun & INTEGER & Version of the SDSS {\tt frames} pipeline used to process the data \\
3  & Field & INTEGER & SDSS field number along a strip \\
4  & Camcol & INTEGER & SDSS camera column \\
5  & Filter & INTEGER & SDSS wave band ($0 = u$, $1 = g$, $2 = r$, $3 = i$, $4 = z$) \\
6  & Object Type & INTEGER & Object classification (3 = Galaxy, 6 = Star) \\
7  & RA & DOUBLE & Right ascension J2000 (deg) \\
8  & Dec & DOUBLE & Declination J2000 (deg) \\
9  & Row & FLOAT & CCD row coordinate (pix) \\
10 & Column & FLOAT & CCD column coordinate (pix) \\
11 & HJD & DOUBLE & Heliocentric Julian date (days) \\
12 & PSF Luptitude & FLOAT & PSF magnitude (lup)$^{*}$ \\
13 & PSF Luptitude Error & FLOAT & Uncertainty on the PSF magnitude (lup)$^{*}$ \\
14 & PSF Flux & FLOAT & PSF flux normalised by the flux from a zero-th magnitude object \\
15 & PSF Flux Error & FLOAT & Uncertainty on the PSF flux \\
16 & Exp Luptitude & FLOAT & Exponential magnitude (lup)$^{*}$ \\
17 & Exp Luptitude Error & FLOAT & Uncertainty on the exponential magnitude (lup)$^{*}$ \\
18 & Exp Flux & FLOAT & Exponential flux normalised by the flux from a zero-th magnitude object \\
19 & Exp Flux Error & FLOAT & Uncertainty on the exponential flux \\
20 & Sky & FLOAT & Sky background brightness (lup arcsec$^{-2}$)$^{*}$ \\
21 & Sky Error & FLOAT & Uncertainty on the sky background brightness (lup arcsec$^{-2}$)$^{*}$ \\
22 & FWHM & FLOAT & Full-width half-maximum of the PSF (arcsec) \\
23 & Photometric Calibration Tag & INTEGER & Flag indicating the photometric calibration status (1 = Calibrated, 0 = Uncalibrated) \\
24 & Photometric Zero Point & FLOAT & Fractional flux offset applied to the flux values \\
25 & Flag 1 & LONG & Object flags (tsObj file tag {\tt FLAGS}) \\
26 & Flag 2 & LONG & More object flags (tsObj file tag {\tt FLAGS2}) \\
27 & Astrometric Calibration Tag & INTEGER & Flag indicating the astrometric calibration status (1 = Calibrated, 0 = Uncalibrated) \\
28 & RA Correction & DOUBLE & Correction applied to right ascension (deg) \\
29 & Dec Correction & DOUBLE & Correction applied to declination (deg) \\
\hline
\end{tabular}
\raggedright
$^{*}$See \citet{lup1999} and \citet{sto2002} for the definition of a luptitude.
\label{tab:lmc}
\end{table*}

A single light-motion curve is stored as an {\tt ASCII} file with a name
constructed from the unweighted mean position of the corresponding object.
The {\tt ASCII} light-motion curve file contains a header line describing the column
meanings, followed by exactly five rows for each epoch (one row for each wave band) in strict time order.
All five wave band measurements are included for completeness, even though it is possible that at any one epoch, up to four
wave band measurements may not satisfy the quality criteria described in Section~2.3. In Table~\ref{tab:lmc} we describe
the columns that make up a light-motion curve from the LMCC.

Figure~\ref{fig:example} shows some clear examples of photometric variability and motion from the LMCC.
Figure~\ref{fig:lc} presents the lightcurve in $r$ (upper points) and $g$ (lower points)
of the large-amplitude long-period variable star SDSS~J220514.58+000845.7, most likely a Mira variable
(\citealt{wat2008}).
Figure~\ref{fig:pm} presents the motion curve of the known ultracool white
dwarf SDSS~J224206.19+004822.7 (\citealt{kil2006}). Both panels illustrate the dramatic increase in temporal sampling 
produced by the start of the SDSS-II Supernova Survey in 2005.

\begin{table*}
\centering
\caption{The list of derived quantities related to photometry that are stored for each light-motion curve in the HLC.
         These quantities are calculated using only light-motion curve entries that satisfy the quality constraints   
         from Section~2.3.}
\begin{tabular}{@{}lll}
\hline
Tag Name In HLC & Type & Description \\
\hline
{\tt LC\_NAME}                & STRING            & Light-motion curve filename \\ 
{\tt IAU\_NAME}               & STRING            & Object name in SDSS Data Release 6 (International Astronomical Union approved format)$^{\dagger}$ \\
{\tt N\_GOOD\_EPOCHS}         & 5 $\times$ INTEGER & Number of good photometric data points \\
{\tt MEAN\_PSFMAG}            & 5 $\times$ FLOAT  & Inverse variance weighted mean of the PSF magnitudes \\
{\tt MEAN\_PSFMAG\_ERR}       & 5 $\times$ FLOAT  & Uncertainty on {\tt MEAN\_PSFMAG} \\   
{\tt MEAN\_EXPMAG}            & 5 $\times$ FLOAT  & Inverse variance weighted mean of the exponential magnitudes \\
{\tt MEAN\_EXPMAG\_ERR}       & 5 $\times$ FLOAT  & Uncertainty on {\tt MEAN\_EXPMAG} \\
{\tt RMS\_PSFMAG}             & 5 $\times$ FLOAT  & Root-mean-square deviation of the PSF magnitudes \\  
{\tt RMS\_EXPMAG}             & 5 $\times$ FLOAT  & Root-mean-square deviation of the exponential magnitudes \\
{\tt CHISQ\_PSFMAG}           & 5 $\times$ FLOAT  & Chi-squared of the PSF magnitudes \\
{\tt CHISQ\_EXPMAG}           & 5 $\times$ FLOAT  & Chi-squared of the exponential magnitudes \\
{\tt N\_GOOD\_EPOCHS\_PSF\_CLIP} & 5 $\times$ INTEGER & Number of good PSF magnitudes after 4$\sigma$-clipping$^{*}$ \\
{\tt N\_GOOD\_EPOCHS\_EXP\_CLIP} & 5 $\times$ INTEGER & Number of good exponential magnitudes after 4$\sigma$-clipping$^{*}$ \\
{\tt MEAN\_PSFMAG\_CLIP}      & 5 $\times$ FLOAT  & 4$\sigma$-clipped inverse variance weighted mean of the PSF magnitudes$^{*}$ \\
{\tt MEAN\_PSFMAG\_ERR\_CLIP} & 5 $\times$ FLOAT  & Uncertainty on {\tt MEAN\_PSFMAG\_CLIP}$^{*}$ \\
{\tt MEAN\_EXPMAG\_CLIP}      & 5 $\times$ FLOAT  & 4$\sigma$-clipped inverse variance weighted mean of the exponential magnitudes$^{*}$ \\
{\tt MEAN\_EXPMAG\_ERR\_CLIP} & 5 $\times$ FLOAT  & Uncertainty on {\tt MEAN\_EXPMAG\_CLIP}$^{*}$ \\
{\tt RMS\_PSFMAG\_CLIP}       & 5 $\times$ FLOAT  & Root-mean-square deviation of the 4$\sigma$-clipped PSF magnitudes$^{*}$ \\
{\tt RMS\_EXPMAG\_CLIP}       & 5 $\times$ FLOAT  & Root-mean-square deviation of the 4$\sigma$-clipped exponential magnitudes$^{*}$ \\
{\tt CHISQ\_PSFMAG\_CLIP}     & 5 $\times$ FLOAT  & Chi-squared of the 4$\sigma$-clipped PSF magnitudes$^{*}$ \\
{\tt CHISQ\_EXPMAG\_CLIP}     & 5 $\times$ FLOAT  & Chi-squared of the 4$\sigma$-clipped exponential magnitudes$^{*}$ \\
{\tt MEAN\_PSFMAG\_ITER}      & 5 $\times$ FLOAT  & Iterated inverse variance weighted mean of the PSF magnitudes$^{*}$ \\
{\tt MEAN\_PSFMAG\_ERR\_ITER} & 5 $\times$ FLOAT  & Uncertainty on {\tt MEAN\_PSFMAG\_ITER}$^{*}$ \\
{\tt MEAN\_EXPMAG\_ITER}      & 5 $\times$ FLOAT  & Iterated inverse variance weighted mean of the exponential magnitudes$^{*}$ \\
{\tt MEAN\_EXPMAG\_ERR\_ITER} & 5 $\times$ FLOAT  & Uncertainty on {\tt MEAN\_EXPMAG\_ITER}$^{*}$ \\
{\tt PERCENTILE\_05\_PSF}     & 5 $\times$ FLOAT  & 5th Percentile of the cumulative distribution of PSF magnitudes \\
{\tt PERCENTILE\_50\_PSF}     & 5 $\times$ FLOAT  & Median of the PSF magnitudes \\
{\tt PERCENTILE\_95\_PSF}     & 5 $\times$ FLOAT  & 95th Percentile of the cumulative distribution of PSF magnitudes \\
{\tt PERCENTILE\_05\_EXP}     & 5 $\times$ FLOAT  & 5th Percentile of the cumulative distribution of exponential magnitudes \\
{\tt PERCENTILE\_50\_EXP}     & 5 $\times$ FLOAT  & Median of the exponential magnitudes \\
{\tt PERCENTILE\_95\_EXP}     & 5 $\times$ FLOAT  & 95th Percentile of the cumulative distribution of exponential magnitudes \\
{\tt TIME\_SPAN}              & FLOAT             & Time span of the light-motion curve (d) \\
{\tt MEAN\_OBJECT\_TYPE}      & FLOAT             & Unweighted mean of the object classification$^{*}$ \\
{\tt MEAN\_CHILD}             & FLOAT             & Unweighted mean of whether the child bit is set or not$^{*}$ \\
{\tt EXTINCTION}              & 5 $\times$ FLOAT  & Galactic extinction (mag)$^{*}$ \\
{\tt ECL\_REDCHISQ\_OUT}      & FLOAT             & Reduced chi-squared out-of-eclipse for the PSF magnitudes$^{*}$ \\
{\tt ECL\_STAT}               & FLOAT             & Eclipse statistic for the PSF magnitudes$^{*}$ \\
{\tt ECL\_EPOCH}              & DOUBLE            & Eclipse epoch as a heliocentric Julian date (d)$^{*}$ \\
{\tt FLARE\_REDCHISQ\_OUT}    & FLOAT             & Reduced chi-squared out-of-flare for the PSF magnitudes$^{*}$ \\
{\tt FLARE\_STAT}             & FLOAT             & Flare statistic for the PSF magnitudes$^{*}$ \\
{\tt FLARE\_EPOCH}            & DOUBLE            & Flare epoch as a heliocentric Julian date (d)$^{*}$ \\
{\tt STETSON\_INDEX\_J\_PSF}  & 5 $\times$ FLOAT  & Stetson J-index for the PSF magnitudes$^{*}$ \\
{\tt STETSON\_INDEX\_J\_EXP}  & 5 $\times$ FLOAT  & Stetson J-index for the exponential magnitudes$^{*}$ \\
{\tt STETSON\_INDEX\_K\_PSF}  & 5 $\times$ FLOAT  & Stetson K-index for the PSF magnitudes$^{*}$ \\
{\tt STETSON\_INDEX\_K\_EXP}  & 5 $\times$ FLOAT  & Stetson K-index for the exponential magnitudes$^{*}$ \\
{\tt STETSON\_INDEX\_L\_PSF}  & 5 $\times$ FLOAT  & Stetson L-index for the PSF magnitudes$^{*}$ \\
{\tt STETSON\_INDEX\_L\_EXP}  & 5 $\times$ FLOAT  & Stetson L-index for the exponential magnitudes$^{*}$ \\
{\tt VIDRIH\_INDEX\_PSF}      & 5 $\times$ FLOAT  & Vidrih index for the PSF magnitudes$^{*}$ \\
{\tt VIDRIH\_INDEX\_EXP}      & 5 $\times$ FLOAT  & Vidrih index for the exponential magnitudes$^{*}$ \\
\hline
$^{*}$See text for more detail. & & \\
$^{\dagger}$May be empty. & & \\
\end{tabular}
\label{tab:hlc1}
\end{table*}

\begin{table*}
\centering
\caption{The list of derived quantities related to astrometry that are stored for each light-motion curve in the HLC.
         These quantities are calculated using only light-motion curve entries that satisfy the quality constraints
         from Section~2.3.}
\begin{tabular}{@{}lll}
\hline
Tag Name In HLC & Type & Description \\
\hline
{\tt RA\_MEAN}                & DOUBLE            & Inverse variance weighted mean of the RA measurements (deg) \\
{\tt RA\_MEAN\_ERR}           & FLOAT             & Uncertainty on {\tt RA\_MEAN} (deg) \\   
{\tt RA\_PM}                  & FLOAT             & Proper motion in the RA coordinate (arcsec yr$^{-1}$) \\
{\tt RA\_PM\_ERR}             & FLOAT             & Uncertainty on {\tt RA\_PM} (arcsec yr$^{-1}$) \\
{\tt RA\_CHISQ\_CON}          & FLOAT             & Chi-squared of the RA measurements for a model that includes only a mean position \\
{\tt RA\_CHISQ\_LIN}          & FLOAT             & Chi-squared of the RA measurements for a model that includes a mean position and a proper motion \\
{\tt RA\_MEAN\_CLIP}          & DOUBLE            & Inverse variance weighted mean of the clipped RA measurements (deg) \\
{\tt RA\_MEAN\_ERR\_CLIP}     & FLOAT             & Uncertainty on {\tt RA\_MEAN\_CLIP} (deg) \\
{\tt RA\_PM\_CLIP}            & FLOAT             & Proper motion in the RA coordinate for the clipped RA measurements (arcsec yr$^{-1}$) \\
{\tt RA\_PM\_ERR\_CLIP}       & FLOAT             & Uncertainty on {\tt RA\_PM\_CLIP} (arcsec yr$^{-1}$) \\
{\tt RA\_CHISQ\_CON\_CLIP}    & FLOAT             & Chi-squared of the clipped RA measurements for a model that includes only a mean position \\
{\tt RA\_CHISQ\_LIN\_CLIP}    & FLOAT             & Chi-squared of the clipped RA measurements for a model that includes a mean position
                                                    and a proper motion \\
{\tt DEC\_MEAN}               & DOUBLE            & Inverse variance weighted mean of the Dec measurements (deg) \\
{\tt DEC\_MEAN\_ERR}          & FLOAT             & Uncertainty on {\tt DEC\_MEAN} (deg) \\
{\tt DEC\_PM}                 & FLOAT             & Proper motion in the Dec coordinate (arcsec yr$^{-1}$) \\
{\tt DEC\_PM\_ERR}            & FLOAT             & Uncertainty on {\tt DEC\_PM} (arcsec yr$^{-1}$) \\
{\tt DEC\_CHISQ\_CON}         & FLOAT             & Chi-squared of the Dec measurements for a model that includes only a mean position \\
{\tt DEC\_CHISQ\_LIN}         & FLOAT             & Chi-squared of the Dec measurements for a model that includes a mean position and a proper motion \\
{\tt DEC\_MEAN\_CLIP}         & DOUBLE            & Inverse variance weighted mean of the clipped Dec measurements (deg) \\
{\tt DEC\_MEAN\_ERR\_CLIP}    & FLOAT             & Uncertainty on {\tt DEC\_MEAN\_CLIP} (deg) \\
{\tt DEC\_PM\_CLIP}           & FLOAT             & Proper motion in the Dec coordinate for the clipped Dec measurements (arcsec yr$^{-1}$) \\
{\tt DEC\_PM\_ERR\_CLIP}      & FLOAT             & Uncertainty on {\tt DEC\_PM\_CLIP} (arcsec yr$^{-1}$) \\
{\tt DEC\_CHISQ\_CON\_CLIP}   & FLOAT             & Chi-squared of the clipped Dec measurements for a model that includes only a mean position \\
{\tt DEC\_CHISQ\_LIN\_CLIP}   & FLOAT             & Chi-squared of the clipped Dec measurements for a model that includes a mean position
                                                    and a proper motion \\
{\tt T0}                      & DOUBLE            & Inverse variance weighted mean of the heliocentric Julian dates using the uncertainties
                                                    on the astrometric measurements (d) \\
{\tt T0\_CLIP}                & DOUBLE            & Inverse variance weighted mean of the heliocentric Julian dates for the
                                                    clipped astrometric measurements (d) \\
{\tt N\_POS\_EPOCHS}          & INTEGER           & Number of good positional measurements \\
{\tt N\_POS\_EPOCHS\_CLIP}    & INTEGER           & Number of good positional measurements after clipping \\
\hline
\end{tabular}
\label{tab:hlc2}
\end{table*}

\section{The Higher-Level Catalogue}

\subsection{Catalogue Description}

The HLC supplies a set of 229 derived quantities for each light-motion curve in
the LMCC. These quantities are aimed at describing the mean magnitudes, photometric variability and astrometric
motion of the objects in the LMCC, and they are calculated using only light-motion curve entries that satisfy the quality constraints
from Section~2.3. Those quantities in the HLC related to photometry are described in Table~\ref{tab:hlc1}, while those
related to astrometry are described in Table~\ref{tab:hlc2}. 

In Table~\ref{tab:hlc1}, if a tag name is associated with a 5-element array, then 
the 5 values represent the described quantity for each of the five SDSS
wave bands in the order $u$, $g$, $r$, $i$ and $z$. When a certain wave band has insufficent ``good'' 
light-motion curve entries to calculate a particular quantity, a value of zero is stored (this also applies to Table~\ref{tab:hlc2}).
For instance, the first value in the array {\tt MEAN\_PSFMAG} is set to zero for any light-motion curves with no ``good'' 
entries for the $u$ band.

All quantities in Table~\ref{tab:hlc1} with {\tt CLIP} at the end of the tag name 
are calculated using a 4$\sigma$-clip algorithm that rejects only the worst outlier at any one
iteration, and terminates when no more outliers are identified. Similarly, all quantities
in Table~\ref{tab:hlc1} with {\tt ITER} at the end of the tag name are calculated using the iterative
procedure described in \citet{ste1996} to dynamically reweight data points based on the size of the residuals from the mean. 
Both these sets of quantities have been designed to be more robust against outliers than a simple inverse variance weighted mean.

The SDSS photometric pipeline performs a morphological star/galaxy separation, the quality of which is intimately
related to seeing and sky brightness. While the accuracy is very good for bright objects, 
there can be confusion for faint objects. The quantity {\tt MEAN\_OBJECT\_TYPE} in Table~\ref{tab:hlc1} 
is an unweighted mean of the SDSS object type classification. Hence it has a value of
3 if the object is classified as a galaxy at all epochs, a value of 6 if the object is classified as a star
at all epochs, and a value between 3 and 6 otherwise. 
The reliability of {\tt MEAN\_OBJECT\_TYPE} for object type classification depends
on the reliability of the SDSS object type classifier and the number of epochs at which the object was observed.

The quantity {\tt MEAN\_CHILD} in Table~\ref{tab:hlc1} is an unweighted mean of whether the child 
object bit is set or not. In other words, this
quantity has a value of 1 if, at all epochs, the object results from the deblending of a parent object, 
a value of 0 if the object was never the result of the deblending of a parent object, 
and a value between 0 and 1 otherwise.

The Galactic extinctions described by {\tt EXTINCTION} in Table~\ref{tab:hlc1} are derived using the maps of dust column density from 
\citet{sch1998}.

For eclipse and flare detection in light curves we include the statistics {\tt ECL\_STAT} and {\tt FLARE\_STAT} (Table~\ref{tab:hlc1}) in the HLC.
In calculating these values, we assume that any eclipse/flare event in a light curve will include only one photometric data point 
since the time elapsed between consecutive scans is at least one day. 
Hence, for each photometric data point (using PSF magnitudes only), we calculate the statistic 
$S^{2}$, based on a matched filter from \citet{bra2005}, and defined by:
\begin{equation}
S^{2} \equiv \frac{ \chi^{2}_{\mbox{\small const}} - \chi^{2}_{\mbox{\small out}} }
            { \left( \displaystyle \chi^{2}_{\mbox{\small out}} /
                 \displaystyle \nu \right) } 
\label{eqn:trastat}
\end{equation}
where $\chi^{2}_{\mbox{\small const}}$ is the chi-squared value of a constant fit for each wave band to the whole light curve, and
$\chi^{2}_{\mbox{\small out}}$ and $\nu$ are the chi-squared and number of degrees freedom, respectively,
of a constant fit for each wave band to the out-of-eclipse/flare light curve. In order to avoid false positives,
we only calculate $S^{2}$ for epochs with photometric data points that are ``good'' in at least two wave bands.
The adopted values of {\tt ECL\_STAT} and {\tt FLARE\_STAT} are then taken to be the largest values 
of $S^{2}$ for photometric data points fainter and brighter, respectively,
than the mean. We also record the corresponding values of $\chi^{2}_{\mbox{\small out}} / \nu$ for {\tt ECL\_STAT} and {\tt FLARE\_STAT} in the 
quantities {\tt ECL\_REDCHISQ\_OUT} and {\tt FLARE\_REDCHISQ\_OUT} respectively, along with the epoch of the putative eclipse/flare event
in {\tt ECL\_EPOCH} and {\tt FLARE\_EPOCH} respectively.

In Table~\ref{tab:hlc1}, the Stetson variability indices J, K and L (\citealt{ste1996}) for both the PSF and exponential
magnitudes are stored in the quantities with tag names starting {\tt STETSON\_INDEX}. We chose the $r$ band as the comparison
wave band and consequently the third element in each of these six quantity arrays is set to zero.

Our final measure of light curve variability is via a quantity called
the Vidrih variability index. RMS magnitude deviations for 
non-variable lightcurves plotted versus mean magnitude $m$ in a given wave band (referred to as an RMS diagram) are 
scattered around a three parameter empirical exponential function $f(m)~=~A~+~B~\,~\exp~(C~\,~(m~-~18))$, 
while the RMS magnitude deviation of variable sources is expected to be noticeably larger. 
We consider two lightcurve samples, that of PSF-like objects (${\tt MEAN\_OBJECT\_TYPE} = 6$) and
non-PSF-like objects (${\tt MEAN\_OBJECT\_TYPE} = 3$). For each lightcurve sample and wave band, we
iteratively fitted $f(m)$ to the corresponding RMS diagrams using a 3$\sigma$-clip algorithm,
employing PSF magnitudes for the star lightcurve sample and exponential magnitudes for the
galaxy lightcurve sample. In each case we also constructed a function $g(m)$ 
describing the standard deviation of the scatter around $f(m)$ via a four-degree polynomial fit to the standard deviation
of the RMS magnitude deviations measured in 0.25 magnitude bins.
Then, for any light curve, the Vidrih index $V$ was calculated as
its RMS magnitude deviation minus $f(m)$, normalized by
$g(m)$, with negative values set to zero, and it is stored in the quantities with tag names starting 
{\tt VIDRIH\_INDEX} (Table~\ref{tab:hlc1}).
Note that due to the way the Vidrih indices are constructed, {\tt VIDRIH\_INDEX\_PSF} is only relevant to
stars and {\tt VIDRIH\_INDEX\_EXP} is only relevant to galaxies.

In Table~\ref{tab:hlc2}, we describe the quantities associated with the object astrometry. Specifically we fit
a proper motion model for celestial coordinates $(\alpha,\delta)$, in degrees, at heliocentric 
Julian date $t$, in days, for each light-motion curve:
\begin{equation}
\alpha(t) = {\tt RA\_MEAN} + \frac{ 7.605 \times 10^{-7} {\tt RA\_PM} }{ \cos( {\tt DEC\_MEAN} ) }  \, ( t - {\tt T0} ) 
\label{eqn:pm_ra}
\end{equation}
\begin{equation}
\delta(t) = {\tt DEC\_MEAN} + 7.605 \times 10^{-7} {\tt DEC\_PM} \, ( t - {\tt T0} )
\label{eqn:pm_dec}
\end{equation}
We set {\tt T0} as the weighted mean of the epochs of observation:
\begin{equation}
{\tt T0} = \frac{ \sum_{j} t_{j} / \sigma_{\alpha}(t_{j})^{2} }{ \sum_{j}  1 / \sigma_{\alpha}(t_{j})^{2} }
\label{eqn:t0}
\end{equation}
where $t_{i}$ represents the set of heliocentric Julian dates for the positional measurements, and $\sigma_{\alpha}(t_{i})$
represents the set of uncertainties on the measured celestial coordinates $\alpha_{i}$ and $\delta_{i}$. 
The uncertainties $\sigma_{\alpha}(t_{i})$ are calculated via Equation~\ref{eqn:sig_coord}.

We solve for the quantities {\tt RA\_MEAN}, {\tt RA\_PM}, {\tt DEC\_MEAN} and {\tt DEC\_PM} by minimising the appropriate chi-squared
using a downhill simplex algorithm. We calculate the uncertainties on
these quantities, namely {\tt RA\_MEAN\_ERR}, {\tt RA\_PM\_ERR}, {\tt DEC\_MEAN\_ERR} and {\tt DEC\_PM\_ERR}, by assuming
that $\cos( {\tt DEC\_MEAN} ) \approx 1$ (a valid assumption for $| \delta | \leq 1\fdg26$) and noting that minimising the
chi-squared for each of the resulting equations is a linear least-squares problem with an analytic solution.
We supply the best-fit chi-squareds via the quantities {\tt RA\_CHISQ\_LIN} and {\tt DEC\_CHISQ\_LIN}, and
we record the number of positional measurements used in the fit in the quantity {\tt N\_POS\_EPOCHS}.
We also supply the chi-squareds {\tt RA\_CHISQ\_CON} and {\tt DEC\_CHISQ\_CON} for a model including only a mean position,
which facilitates the calculation of the following delta chi-squared:
\begin{align}
\Delta \chi^{2}_{\mu} = \,\, & {\tt RA\_CHISQ\_CON} - {\tt RA\_CHISQ\_LIN} \notag \\
                       & + {\tt DEC\_CHISQ\_CON} - {\tt DEC\_CHISQ\_LIN} \notag \\
\label{eqn:pm_s2n}
\end{align}
The statistic $\Delta \chi^{2}_{\mu}$ may be used to calculate the significance of a proper motion measurement by
noting that it follows a chi-square distribution with two degrees of freedom.

The whole fitting process is also iterated using a clipping algorithm that rejects the worst outlier at any one
iteration, and terminates when the change in the fitted proper motion is smaller than 3~mas~yr$^{-1}$. 
Consequently, all the derived astrometric quantities 
described so far have a corresponding quantity calculated for the clipped positional measurements and stored
in the HLC (all quantities listed in Table~\ref{tab:hlc2} with a tag name ending {\tt CLIP}). The astrometric quantities
derived from the clipped astrometric data are more robust than those derived from the unclipped astrometric data.
We strongly advise that any statistical studies undertaken with the astrometry in the HLC 
should only use the clipped quantities.

Finally, we have attempted to detect the parallax signal for nearby stars by fitting a parallax model to each light-motion curve.
However, our investigation has made it clear that
the distribution of observations during the same few months each year along with the
positional accuracy of the astrometric measurements are not enough to enable the detection of a clean parallax signal.

\subsection{Catalogue Format}

The HLC is stored in eight FITS binary tables, one for each hour in RA from 20$^{\mbox{\small h}}$ to
4$^{\mbox{\small h}}$. Each record in the HLC corresponds to a single light-motion curve and stores
229 derived quantities with tag names listed and described in Tables~\ref{tab:hlc1} and \ref{tab:hlc2}.
The FITS files ($\sim$1.9~Gb compressed) may be obtained by web download from 
{\it http://das.sdss.org/value\_added/stripe\_82\_variability/SDSS\_82\_public/}.

The IDL Astronomy User's 
Library\footnote{The IDL Astronomy User's Library is web-hosted at http://idlastro.gsfc.nasa.gov/ and maintained by
Wayne Landsman at the Goddard Space Flight Centre.} function ``mrdfits'' is a convenient way to read-in the
HLC, storing the FITS binary table automatically in an IDL structure. The IDL ``where'' function then becomes
a very powerful tool for accessing subsets of data with ease.

\section{External Comparisons}

\subsection{Photometry}

\begin{figure*}
\epsfig{file=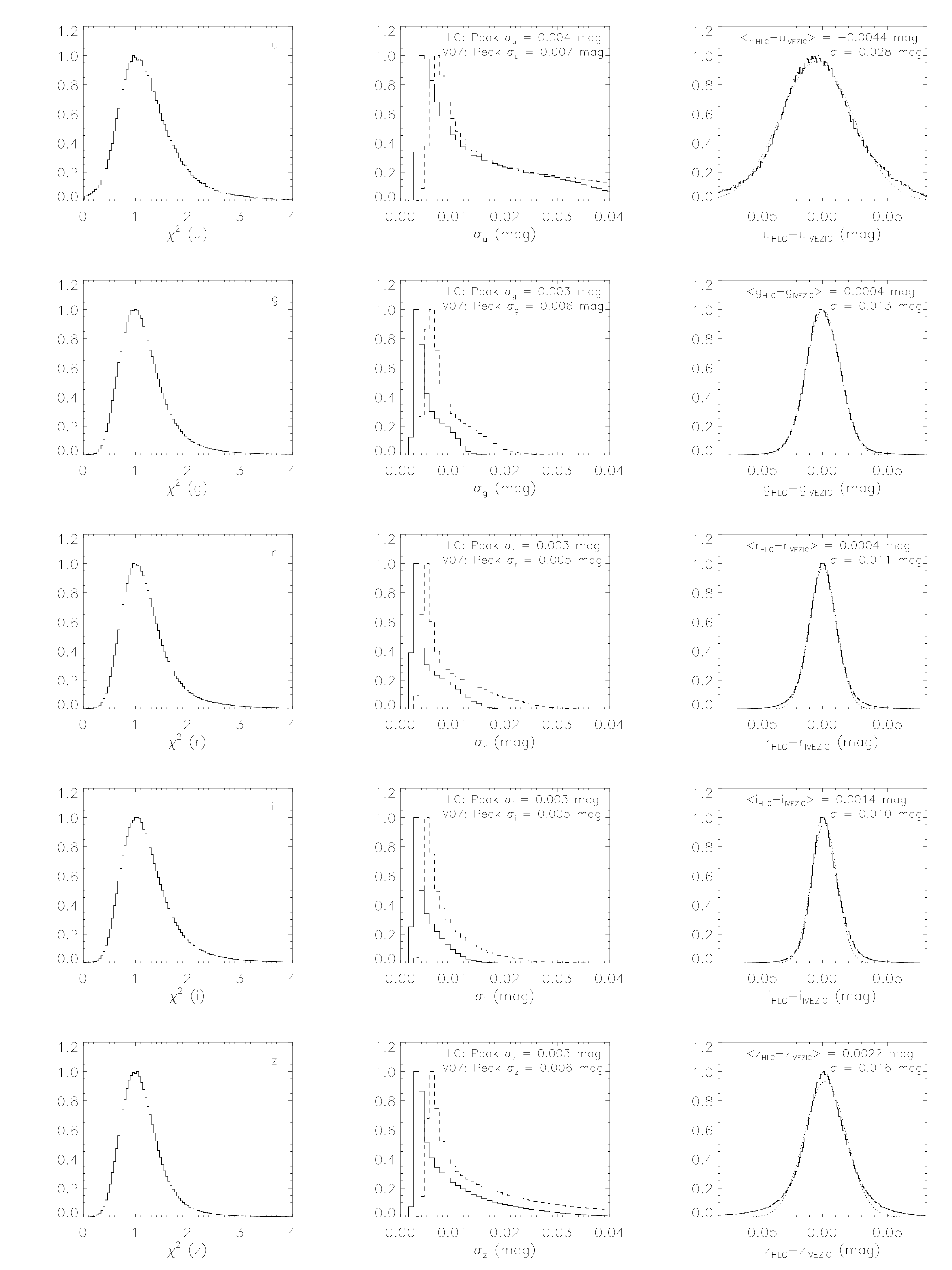,angle=0.0,width=0.92\linewidth}
\caption{Each row of panels corresponds to a different wave band and the order employed is $u$, $g$, $r$, $i$ and $z$ from
         the top row to the bottom row. The LMCC objects used in the plots are matches with the IV07 standard star catalogue
         and have at least 3 good epochs with mean PSF magnitudes brighter than 21.5~mag in the $u$, $g$, $r$ and $i$ bands,
         and brighter than 20.5~mag in the $z$ band. {\bf Left hand column:} Normalised histograms of $\chi^{2}$ per degree
         of freedom for the mean PSF magnitudes in the HLC. {\bf Middle column:} Normalised histograms of the
         uncertainties on the mean PSF magnitudes in the HLC (solid lines) and on the mean magnitudes from the IV07 catalogue
         (dashed lines). {\bf Right hand column:} Normalised histograms of the difference between the HLC mean PSF magnitudes
         and the IV07 catalogue mean magnitudes, along with a fitted Gaussian (dotted line).
         \label{fig:comp_ive}}
\end{figure*}

The most relevant comparison of our catalogue photometry can be made by comparing our results to those
of \citet{ive2007} (from now on IV07) who use 58 pre-2005 SDSS-I imaging runs of Stripe 82, observed under mostly
photometric conditions, to construct a standard star catalogue.
IV07 match unsaturated point sources satisfying high signal-to-noise criteria (photometric uncertainties below
0.05~mag) between runs and choose sources with at least 4 epochs. Those objects that are identified as
non-varying, by requiring a $\chi^{2}$ per degree of freedom for the mean magnitudes of less than 3 in the $g$, $r$ and $i$ bands, 
are chosen as candidate standard stars. 
The standard star catalogue is then internally recalibrated, correcting for flat-field errors
and time-variable extinction (for example, due to clouds). 

The main differences between the LMCC and the IV07
catalogue are that we use a different photometric recalibration method, but based on the same idea of 
spatially dependent photometric zeropoints, and that we include 
the generally non-photometric observations from the SDSS-II supernova runs.
Not surprisingly, for the 886396 objects from the IV07 catalogue in the overlap area with the LMCC, 
we find unambiguous matches in the LMCC for 886191 objects using a match radius of 0.7\arcsec, where the missing fraction
of $\sim$0.02\% can be explained by the different criteria used to construct the two catalogues. 
Of the 886191 matching objects in the LMCC, 219627, 606658, 879766, 886028 and 856456 
objects have at least 3 good epochs in the $u$, $g$, $r$, $i$ and $z$ bands, respectively, and mean PSF magnitudes brighter than 
21.5, 21.5, 21.5, 21.5 and 20.5~mag, respectively. It is these objects that we
use when constructing the histograms in Figure~\ref{fig:comp_ive}.

In Figure~\ref{fig:comp_ive}, each row of panels corresponds to a different wave band and the order employed is 
$u$, $g$, $r$, $i$ and $z$ from the top row to the bottom row. The left hand column of plots are normalised
histograms of $\chi^{2}$ per degree of freedom for the mean PSF magnitudes in the HLC, and they 
serve to confirm that the photometric uncertainties
in the LMCC (inherited from the SDSS pipelines, with some adjustment during photometric recalibration) are correct, since the
histograms peak at a value of $\chi^{2}~/~(N-1)~\sim$~1. 

The middle column of panels in Figure~\ref{fig:comp_ive} are normalised histograms of the uncertainties on the 
mean PSF magnitudes in the HLC (solid lines) and on the mean magnitudes from the IV07 catalogue (dashed lines). It is clear that the mean magnitudes
quoted in the HLC are approximately twice as precise as those from IV07, which is due to the greater number of epochs included
in the LMCC. However, systematic errors are still likely to dominate the mean magnitudes at the $\sim$1\% level,
which corresponds to the level of the photometric recalibrations and to the systematic error
introduced by the slightly different band passes of each column of detectors in the SDSS camera. We do not
correct the LMCC magnitudes for the different bandpasses in contrast to IV07.

\begin{figure*}
\epsfig{file=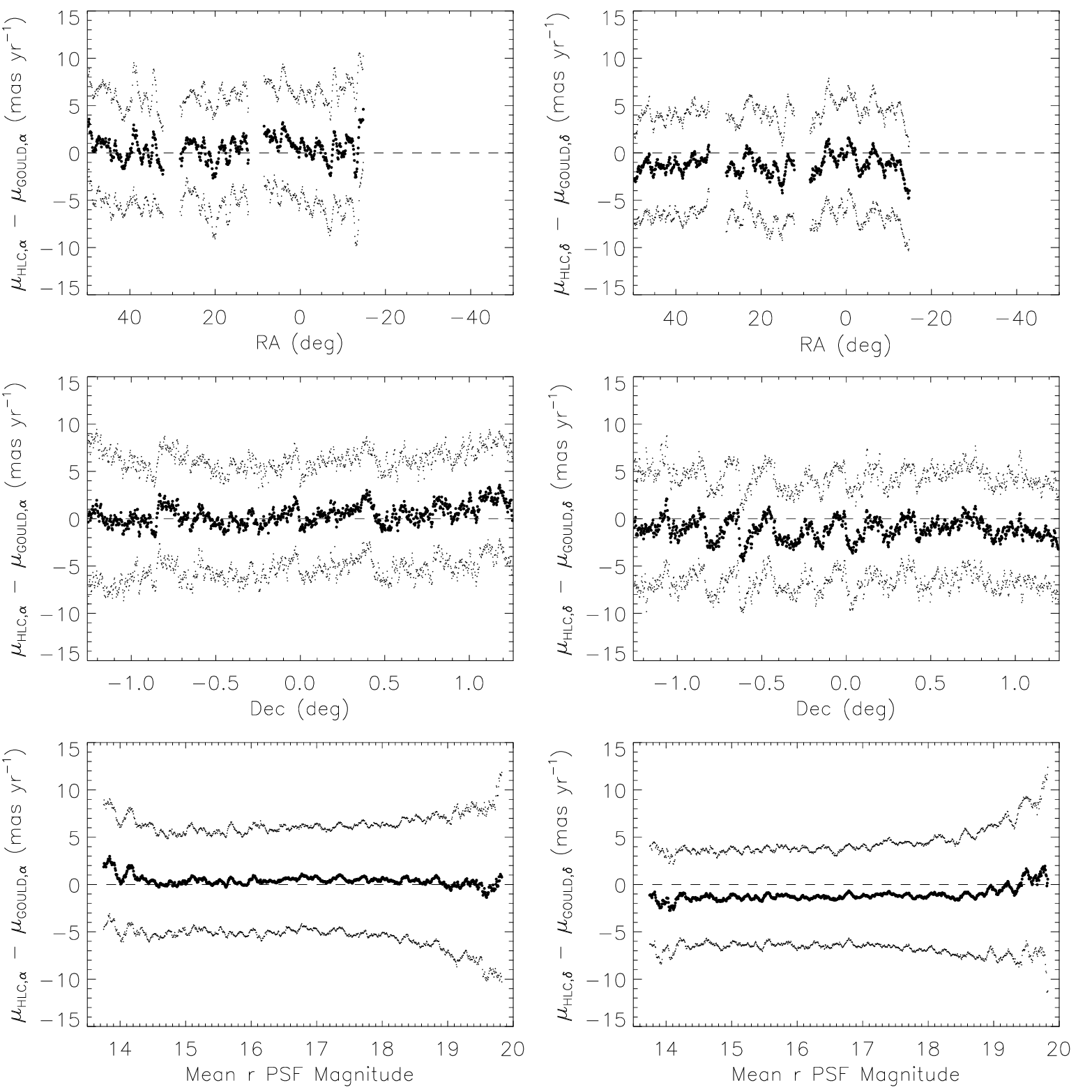,angle=0.0,width=\linewidth}
\caption{Plots of the running 3$\sigma$-clipped mean difference between the HLC proper motions and those of
         GK04 (middle points) as a function of RA using a 1$^{\circ}$ window (upper panels),
         as a function of Dec using a 0\fdg01 window (middle panels),
         and as a function of mean $r$ PSF magnitude using a 0.1~mag window (lower panels).
         The left hand panels correspond to proper motion in RA and the right hand panels correspond to proper motion in Dec.
         The upper and lower sets of smaller points in each panel
         represent the mean difference plus or minus the running 3$\sigma$-clipped standard deviation, respectively.
         \label{fig:comp_usno}}
\end{figure*}

The right hand column of panels in Figure~\ref{fig:comp_ive} are normalised histograms of the difference between the HLC
mean PSF magnitudes and the IV07 catalogue mean magnitudes. A Gaussian has been fitted to each histogram (dotted lines), and the 
fitted centre and sigma are quoted in each panel. The histograms are centred around zero at the millimagnitude level, except for
the $u$ band where the IV07 photometry is slightly offset from the HLC photometry by $\sim$4~mmag. The histograms have sigmas of
28, 13, 11, 10 and 16~mmag in $u$, $g$, $r$, $i$ and $z$, respectively, which are consistent with the scatter in the 
photometric zeropoints of the standard SDSS runs derived in Section~2.2, and with the scatter in the similar internal photometric 
recalibrations of IV07.

The data set used to construct the IV07 standard star catalogue is a subset of the data used to construct the LMCC, and therefore we can
use our superior temporal coverage to identify standard star candidates that are actually photometrically variable.
\citet{wat2008} use the HLC to systematically identify variable stars in Stripe 82, 
and we choose to follow all but one of their cuts on the HLC quantities.
Of the 886191 objects from the IV07 catalogue that have matching objects in the LMCC, 878172 have at least 11
good epochs in the $g$ band, allowing for the calculation of reliable variability and object type indicators. 
The requirement from \citet{wat2008} that an LMCC object has a 
mean object type greater than 5.5 ensures that chosen variables are PSF-like (stars).
However, this implies that 16529 objects are probably non-PSF-like, corresponding to a fraction of $\sim$1.9\%.
Hence, we do not apply this requirement, and instead we directly apply the remaining requirements that a variable object
should have a $\chi^{2}$ per degree of freedom greater than 3 for both the $g$
and $r$ band mean PSF magnitudes, and a Stetson L index greater than 1 for the $g$ band. 
We find that just 570 IV07 standard stars, from 878172, are variable, corresponding to a fraction of $\sim$0.065\%. This confirms that
even with the addition of many more photometric data, the IV07 standard stars are in general still found to be constant at the $\sim$0.01~mag
level. In considering the full IV07 catalogue with 1006849 candidate standard stars, we suspect that $\sim$650 are actually variables.

\begin{figure*}
\epsfig{file=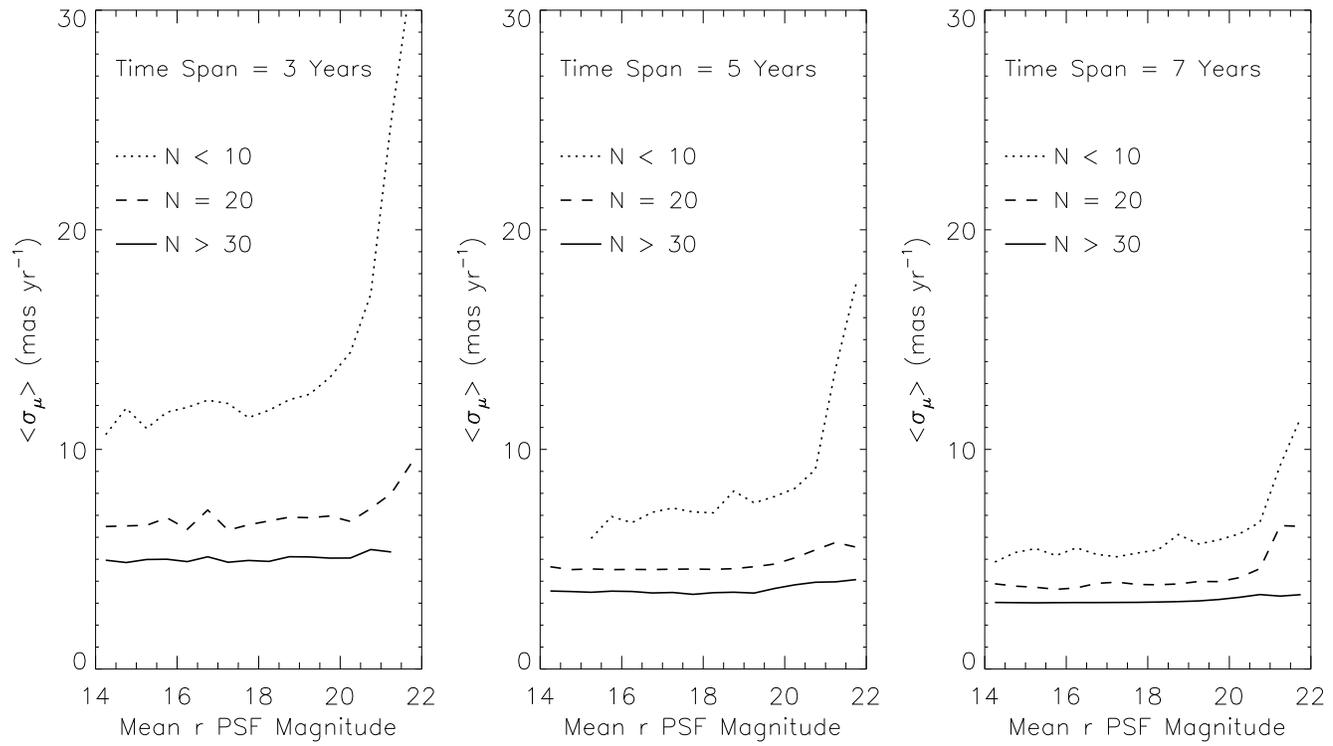,angle=0.0,width=\linewidth}
\caption{Plot of mean proper motion uncertainty in the HLC as a function of mean $r$ PSF magnitude for
         PSF-like objects (stars). The left, middle and right hand panels correspond to light-motion
         curve time spans of 3, 5, and 7 years, respectively. In each panel, the dotted, dashed and continuous
         curves correspond to $\le$10, $\approx$20 and $\ge$30 epochs, respectively.
         \label{fig:pmerr}}
\end{figure*}

\subsection{Astrometry}

\citet{gou2004} (from now on GK04) carefully combine SDSS Data Release~1 (SDSS~DR1) and USNO-B proper motions 
to produce a catalogue of 390476 objects with proper motions $\mu \geq 20$~mas~yr$^{-1}$ and
magnitudes $r \leq 20$~mag. SDSS~DR1 proper motions are based on matches of SDSS to USNO-A2.0
(\citealt{mon1998}), which suffer from mismatches causing spurious high proper-motion
objects, and systematic trends in proper motion. However, by cross-correlating
SDSS DR1 and USNO-B, GK04 successfully removed the vast majority of spurious proper-motion stars.
Furthermore, by considering the set of spectroscopically confirmed quasars in SDSS~DR1, GK04 
calibrate out the position-dependent astrometric biases, using a very similar method to our recalibration of 
SDSS astrometry presented in Section~2.4. 

We compare our proper motions in the HLC to those derived by GK04 for the 30546 stars from their catalogue 
with an unambiguous positional match in the HLC using a match radius of 0.7\arcsec.
In Figure~\ref{fig:comp_usno}, we plot the running 3$\sigma$-clipped mean difference between our HLC proper motions and those of 
GK04 (middle points) as a function of RA using a 1$^{\circ}$ window (upper panels), as a function of Dec using
a 0\fdg01 window (middle panels), and as a function of mean $r$ PSF magnitude using a 0.1~mag window (lower panels).
The mean difference is only calculated if there are at least 100 stars in the sliding window.
The left hand panels correspond to proper motion in RA and the right hand panels correspond to proper motion in Dec.
We also plot the mean difference plus or minus the running 3$\sigma$-clipped standard deviation as the upper
and lower sets of smaller points, respectively, in each panel.

Figure~\ref{fig:comp_usno} illustrates that the systematic differences between the HLC proper motions and those of GK04
are at a very small level, generally $\la$2~mas~yr$^{-1}$. We note one clear systematic trend that the GK04 Dec proper motions are offset from the
HLC Dec proper motions by $\la$2~mas~yr$^{-1}$, which is especially visible in the bottom right hand panel of Figure~\ref{fig:comp_usno}.
We are currently unable to identify unambiguously the origin of the small systematic offset.
The scatter in the proper motion differences (represented by the upper and lower sets of smaller points in each panel) is consistent with
the stated proper motion uncertainties of $\sim$3.9~mas~yr$^{-1}$ in GK04, and $\la$5~mas~yr$^{-1}$ for these particular stars in the HLC.

Each light-motion curve in the LMCC has a different temporal coverage and number of epochs, a situation which is highlighted in Figure~\ref{fig:tcov}.
Consequently the uncertainties on the HLC proper motions exhibit a very inhomogeneous spatial distribution, and selecting proper motion objects
using cuts on proper motion uncertainty results in a very inhomogeneous sample of objects. In Figure~\ref{fig:pmerr}, we present the mean
proper motion uncertainty in the HLC as a function of mean $r$ PSF magnitude, number of epochs and time span of a light-motion curve for
PSF-like objects ({\tt MEAN\_OBJECT\_TYPE~=~6}).
The three panels from left to right correspond to different light-motion curve time spans of 3, 5 and 7 years, respectively, while in each 
panel three curves are plotted, dotted, dashed and continuous, corresponding to $\le$10, $\approx$20 and $\ge$30 epochs, respectively.
The curves in each panel show the mean proper motion uncertainty in the HLC as a function of mean $r$ PSF magnitude.
It is worth noting that $\sim$76\% of objects in the LMCC have at least 20 epochs, and that $\sim$37\% have at least 20 epochs with a time
span of greater than 4 years. We find that 312819 objects in the LMCC (or $\sim$8\%) have proper motions $\mu$ with $\mu / \sigma_{\mu} > 5$.

\section{Conclusions}

The Light-Motion Curve Catalogue (LMCC) contains almost 4~million light-motion curves
for stars and galaxies covering $\sim$249~deg$^{2}$ in the SDSS Stripe 82.
A light-motion curve provides wave band and time dependent
photometric and astrometric quantities, where $\sim$76 per cent of light-motion curves in the LMCC have 
at least 20 epochs of measurements. The LMCC is complete to magnitude 21.5
in $u$, $g$, $r$ and $i$, and to magnitude 20.5 in $z$, making it the deepest large-area catalogue of its kind. 
The photometric RMS accuracy for stars is $\sim$20~mmag at $r\sim$18~mag and for galaxies it is $\sim$30~mmag
at $r\sim$18~mag. In both the RA and Dec coordinates, an RMS accuracy of $\sim$32~mas and $\sim$35~mas at
$r\sim$18~mag is achieved for pre-2005 data for stars and galaxies, respectively, and an RMS accuracy of
$\sim$35~mas and $\sim$46~mas at $r\sim$18~mag is achieved for 2005 data for stars and galaxies, respectively.

The LMCC is thus an ideal tool for studying the variable sky, and in order to aid in its exploitation, 
we have created the Higher-Level Catalogue (HLC). The HLC consists of 229 derived photometric and astrometric quantities 
for each light-motion curve, and it is stored in a set of FITS binary tables, a format that is widely used by the astronomical community. 
The photometry presented in the HLC is fully consistent with the IV07 standard star catalogue and,
since it is based on many more epochs, the random uncertainties in the mean magnitudes are smaller. 
Reassuringly, we show that only a very small percentage of standard stars ($\sim$0.065\% or
$\sim$650 objects) from IV07 are actually photometrically variable.
The HLC proper motions of 30546 stars are consistent to within uncertainties with those derived by \citet{gou2004} 
by combining SDSS~DR1 and USNO-B proper motions.

The power in using the HLC is well illustrated by the work of \citet{vid2007} who construct a reduced proper motion diagram for
Stripe 82 in order to find rare white dwarf populations, and consequently they identify 8 new candidate ultracool white dwarfs
and 10 new candidate halo white dwarfs. Also, \citet{bec2008} report the discovery of an eclipsing M-dwarf binary system 
2MASS~J01542930+0053266, having employed the corresponding light-motion curve along with radial velocity measurements
to determine the masses and radii of the stellar components.
We therefore encourage the astronomical community to actively use and explore these 
catalogues that are so ample in their content.

\section*{Acknowledgements}

The calculations presented in this work were done using IDL programs and the Sun
Grid Engine distributed computing software. IDL is provided, under license, by 
Research Systems Inc. We thank Robert Lupton for his advice on the photometric 
quality flags. We also appreciate the useful comments on astrometry
given by Siegfried R\"oser. D.M.~Bramich is grateful to the Particle 
Physics and Astronomy Research Council (PPARC) for financial support. S.~Vidrih
acknowledges the financial support of the European Space Agency. L.~Wyrzykowski 
was supported by the European Community's Sixth Framework Marie Curie Research 
Training Network Programme, Contract No. MRTN-CT-2004-505183 ``ANGLES''.
 
Funding for the SDSS and SDSS-II has been provided by the Alfred P. Sloan Foundation, 
the Participating Institutions, the National Science Foundation, the U.S. Department of 
Energy, the National Aeronautics and Space Administration, the Japanese Monbukagakusho, 
the Max Planck Society, and the Higher Education Funding Council for England. The SDSS 
Web Site is http://www.sdss.org/.

The SDSS is managed by the Astrophysical Research Consortium for the Participating Institutions. 
The Participating Institutions are the American Museum of Natural History, 
Astrophysical Institute Potsdam, University of Basel, University of Cambridge, 
Case Western Reserve University, University of Chicago, Drexel University, Fermilab, the 
Institute for Advanced Study, the Japan Participation Group, Johns Hopkins University, 
the Joint Institute for Nuclear Astrophysics, the Kavli Institute for Particle 
Astrophysics and Cosmology, the Korean Scientist Group, the Chinese Academy of Sciences (LAMOST), 
Los Alamos National Laboratory, the Max-Planck-Institute for Astronomy 
(MPIA), the Max-Planck-Institute for Astrophysics (MPA), New Mexico State University, 
Ohio State University, University of Pittsburgh, University of Portsmouth, Princeton 
University, the United States Naval Observatory, and the University of Washington.

\label{lastpage}

\end{document}